\documentclass[twocolumn,superscriptaddress,nobalancelastpage,10pt,pra,letter]{revtex4}
\usepackage{amsmath}
\usepackage{amsthm}
\usepackage{amssymb}
\usepackage{amsfonts}
\usepackage{graphicx}
\usepackage{wasysym}
\usepackage{mathrsfs}
\usepackage{yfonts}
\usepackage{bbold}
\usepackage{verbatim}
\usepackage{subfigure}
\usepackage{color}
\usepackage{setspace}
\usepackage{multirow}
\usepackage{float}

\newcommand{\ket}[1]{| #1 \rangle}
\newcommand{\bra}[1]{\langle #1 |}

\newcommand{\dm}{\widehat{\rho}}
\newcommand{\dmt}{\widehat{\widetilde{\rho}}}

\newcommand{\I}{\mathscr{I}}

\newcommand{\V}{\mathcal{V}}

\makeatletter
\newcommand*{\rom}[1]{\expandafter\@slowromancap\romannumeral #1@}
\makeatother

\makeatletter
\newcommand{\roml}[1]{\lowercase\expandafter{\romannumeral #1\relax}}
\makeatother

\DeclareMathAlphabet{\mathpzc}{OT1}{pzc}{m}{it}

\begin{document}

	\title{
    Complete Measurement of Two-Photon Density Matrix by Single-Photon Detection}
	
	\author{Salini Rajeev}
	\affiliation{Department of Physics, 145 Physical Sciences Bldg., Oklahoma State University, Stillwater, OK 74078, USA.}
	
	\author{Mayukh Lahiri}
	\email{mlahiri@okstate.edu} \affiliation{Department of Physics, 145 Physical Sciences Bldg., Oklahoma State University, Stillwater, OK 74078, USA.}

\begin{abstract}
The reconstruction of density matrices from measurement data (quantum state tomography) is the most comprehensive method for assessing the accuracy and performance of quantum devices. Existing methods to reconstruct two-photon density matrices require the detection of both photons unless a priori information is available. Based on the concept of quantum-induced coherence by path identity, we present an approach to quantum state tomography that circumvents the requirement of detecting both photons. We show that an arbitrary two-qubit density matrix, which can contain up to fifteen free parameters, can be fully reconstructed from single-photon measurement data without any postselection and a priori information. In addition to advancing an alternative approach to quantum state measurement problems, our results also have notable practical implications. A practical challenge in quantum state measurement arises from the fact that effective single-photon detectors are not readily accessible for a wide spectral range. Our method, which eliminates the need for coincidence measurements, enables quantum state tomography in the case where one of the two photons is challenging or impossible to detect. It therefore opens the door to measuring quantum states hitherto inaccessible.
\end{abstract}
 
\maketitle


\section{Introduction}\label{sec:intro}

Measurement of quantum states is a fundamental topic of quantum mechanics \cite{von2018mathematical}. A complete reconstruction of the quantum state (density matrix), which is also called quantum state tomography (QST), has been widely studied for various photonic and atomic systems (see, for example, \cite{Vogel1989Wigner,royer1989measurement,Raymer1993tomo,1996-Wineland-tomo_PRL,breitenbach1997measurement,Banaszek1999Maximum,James2001,agnew2011tomography,bayraktar2016quantum}). QST is of central importance to atomic, molecular, and optical (AMO) physics, and quantum information science. QST of atomic systems can be carried out without performing direct measurement on the system (e.g., by detecting fluorescence light \cite{1996-Wineland-tomo_PRL}). However, for photonic systems, all existing methods require performing direct measurement on the entire system. Consider, for example, a two-photon quantum state. If the density matrix is to be reconstructed, both photons are required to be detected \cite{James2001,agnew2011tomography}. Even for problems that do not require retrieving the complete information of a quantum state \textemdash ~for example, entanglement measurement problems \textemdash ~detecting both photons is usually necessary if the two-photon state is not pure (see, for example, \cite{walborn2006,sahoo2020quantum,bhattacharjee2022measurement,Pires2009,Just2013,sharapova2015schmidt}). A fundamental question in this context is: \emph{how much information about an arbitrary two-photon state can be obtained by detecting only one photon}? 
\par
Recent developments in quantum interferometry, particularly regarding the phenomenon of ``induced coherence without induced emission'' \cite{zou1991induced,wang1991induced,chekhova2016nonlinear,hochrainer2022quantum}, provide new insights into this question. This type of interferometry has been used to characterize entanglement in a small class of two-photon mixed states by detecting only one of the photons \cite{lahiri2021characterizing,lemos2023one,zhan2021determining,rajeev2023single}; the states currently covered are mixed states derived from Bell states in two \cite{lahiri2021characterizing,lemos2023one} and higher dimensions \cite{zhan2021determining}, and from two-dimensional (two-qubit) Werner states \cite{rajeev2023single}. Here, we take a significant step by showing that it is possible to completely measure the density matrix of any two-qubit photonic state by detecting only one of the two qubits. 
\par
Specifically, we present a comprehensive theory of two-qubit quantum state tomography based on the phenomenon of ``induced coherence without induced emission'' \cite{zou1991induced}, also known as ``interference by path identity'' \cite{hochrainer2022quantum}. Unlike standard state tomography techniques, our method does not require coincidence measurement or postselection; it only requires measuring single-photon detection probability (counting rate). Therefore, from a practical standpoint, our approach is advantageous when a suitable detector is unavailable for one of the two photons.
\par
This article is organized as follows: In Sec.~\ref{sec:principle}, we discuss the principle on which our method relies. Section~\ref{sec:analytical-descrpn} contains the detailed theoretical analysis; this section is divided into seven subsections: In subsection \ref{subsec:gen-q-state}, we represent an arbitrary two-qubit density matrix in a form suitable for our analysis. In subsection~\ref{subsec:state-para}, we discuss the connection between state parameters and state creation. In subsection~\ref{subsec:outline}, we give an outline of the tomography scheme, including a description of the interferometer employed. Subsections \ref{subsec:total-quantum-state}, \ref{subsec:alignment-conditions}, and \ref{subsec:phtn-detect-probty} show how to obtain analytical expressions for single-photon interference patterns starting from quantum states generated by individual sources in the interferometer. In subsection \ref{subsec:state-reconstruction}, we show how to obtain all density matrix elements from the single-photon interference patterns. In Sec.~\ref{sec:illustration}, we numerically illustrate our results choosing a two-qubit mixed state derived from a Bell state. In Sec.~\ref{sec:loss-main}, we discuss how to treat the key anticipated experimental imperfections (Appendix \ref{app:gen-treatment-loss} presents the corresponding analysis in detail). Finally, in Sec.~\ref{sec:discuss}, we briefly discuss our results and present an outlook.  

\section{Principle}\label{sec:principle}
\begin{figure*} \centering
	\includegraphics[width=\linewidth]{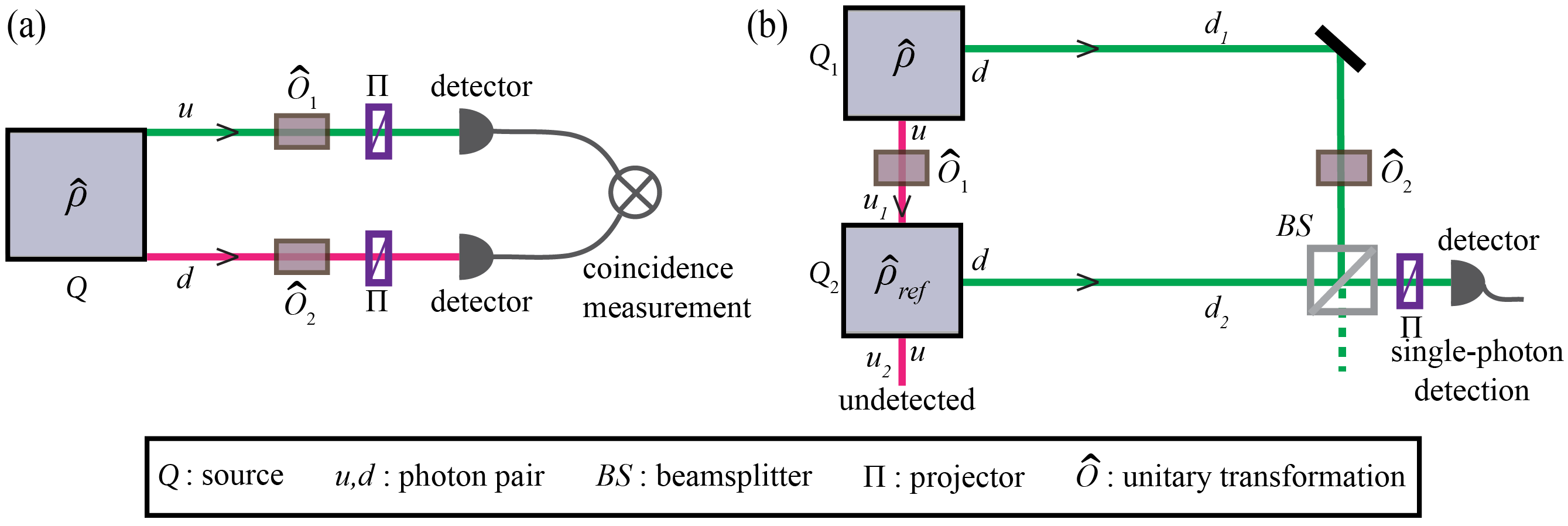}
	\caption{Principle and comparison with traditional quantum state tomography. (a) In traditional tomography of an unknown two-qubit photonic state ($\dm$), both photons ($d$, $u$) must be detected via coincidence or an equivalent measurement technique. The traditional approach is not interferometric. (b) Our method is interferometric. It requires the detection of only one of the two photons ($d$), while the other photon ($u$) remains undetected. The setup contains two sources, $Q_1$ and $Q_2$. Source $Q_1$ emits the twin photons ($d$,$u$) in the state $\hat{\rho}$ that we want to measure, while $Q_2$ emits them in a known reference state $\hat{\rho}_{ref}$. The photon $d$ ($u$) propagates as beams $d_1$ and $d_2$ ($u_1$ and $u_2$) after emerging from $Q_1$ and $Q_2$. Beams $d_1$ and $d_2$ are superposed by a beamsplitter and one of the outputs is sent to a detector. Each $d$-photon is projected onto a suitable state before detection. A unitary transformation, $\hat{O}_2$, is applied to the $d$-photon in beam $d_2$. The $u$-photon undergoes a unitary transformation $\hat{O}_1$ while in beam $u_1$. Beam $u_1$ is sent through $Q_2$ and aligned with $u_2$. Such an alignment (path identity) makes beams $d_1$ and $d_2$ mutually coherent and, consequently, single-photon interference is observed at the detector. The unknown state $\dm$ is reconstructed from the single-photon interference data without any postselection.} \label{fig:scheme}
\end{figure*} 
Traditional quantum tomography of two-qubit photonic states relies on the detection of both photons via coincidence measurements or an equivalent measurement technique (Fig.~\ref{fig:scheme}a). In contrast, our proposed tomography scheme relies on a quantum interferometric approach in which only one of the two photons is detected (Fig.~\ref{fig:scheme}b). In our case, only single-photon counting rate (quantum mechanical analogue of intensity) is measured and no post-selection is performed. This type of interferometry was discovered by Zou, Wang, and Mandel in 1991 \cite{zou1991induced,wang1991induced} and has also found applications in quantum imaging \cite{lemos2014quantum}, sensing \cite{kalashnikov2016infrared,qian2023quantum}, and remote sensing \cite{dalvit2024quantum}.
\par
We denote the two photons constituting a pair by $d$ and $u$; photon $d$ is detected and photon $u$ remains undetected in our scheme. We construct an interferometer containing two sources, $Q_1$ and $Q_2$, each of which can generate a photon pair (Fig.~\ref{fig:scheme}b). Source $Q_1$ emits the photon pair ($d,u$) in a generally mixed state, $\dm$, which is to be reconstructed. No prior assumption is made on state $\dm$. The other source $Q_2$ emits the same photon pair $(d,u)$ in a known reference state $\dm_{ref}$. We assume that the probability of finding states with a photon number exceeding two, as well as contributions from stimulated emissions, is negligible. In experiments designed with nonlinear SPDC crystals, this condition is satisfied when the crystals are weakly pumped \cite{zou1991induced,wang1991induced,wiseman2000induced,liu2009investigation,kolobov2017controlling,lahiri2019nonclassicality,barreto2022quantum}.
\par
Sources $Q_1$ and $Q_2$ emit photon $d$ into beams $d_1$ and $d_2$, respectively. These two beams are superposed by a beam splitter ($BS$) and one of the outputs of $BS$ is sent to a detector where single-photon counting rate (intensity) is measured (Fig.~\ref{fig:scheme}b). Photon $d$ is projected onto a suitable state before its detection.
\par
Photon $u$ is emitted into beams $u_1$ and $u_2$ by sources $Q_1$ and $Q_2$, respectively. Beam $u_1$ is sent through $Q_2$ and aligned with beam $u_2$ as shown in Fig.~\ref{fig:scheme}b. Such an alignment (originally proposed by Ou) makes paths of photon $u$ emerging from the two sources identical, and therefore is also called path identity \cite{hochrainer2022quantum}. Path identity makes it impossible to know from which source $d$-photon has arrived at the detector (unless the information is available via another means). Due to the unavailability of the which-path information, single-photon interference can be observed at the detector. \emph{A remarkable fact is that the interference pattern can be manipulated using photon $u$, which is not detected to obtain the interference pattern.} Following the terminology of Ref.~\cite{hochrainer2022quantum}, we will call such an interference as \emph{interference by path identity}. 
\par
We apply a unitary transformation, $\hat{O}_1$, to $u$-photon in its way from $Q_1$ to $Q_2$ and another unitary transformation, $\hat{O}_2$, to $d$-photon in beam $d_2$. A set of single-photon interference patterns are obtained for various choices of these unitary transformations. We show below that the information of the unknown two-qubit state ($\dm$) appears in these interference patterns. We also show that the unknown state can be fully reconstructed using the single-photon interference patterns without any prior information by appropriately choosing $\dm_{ref}$, $\hat{O}_1$ and $\hat{O}_2$.

\section{Theoretical Analysis}\label{sec:analytical-descrpn}

\subsection{An arbitrary two-qubit state}\label{subsec:gen-q-state}
We work with a two-photon mixed state in the polarization basis. We use $H$ and $V$ to represent the horizontal and vertical polarization respectively. In the computational basis $\{\ket{H_uH_d}, \ket{H_uV_d}, \ket{V_uH_d}, \ket{V_uV_d}\}$, where $d$ and $u$ denote the two photons, an arbitrary two-qubit mixed state can be represented by the following density operator (density matrix) without any loss of generality [for matrix form, see Appendix \ref{app:matrix-dm}, Eq.~(\ref{q-state-matrix})]: 
\begin{align}\label{q-state}
\dm=\sum_{\mu,\nu}^{H,V}\sum_{\mu',\nu'}^{H,V}\sqrt{I_{\mu\nu}I_{\mu'\nu'}}J_{\mu\nu}^{\mu'\nu'}e^{i\phi_{\mu\nu}^{\mu'\nu'}}\ket{\mu_u \nu_d}\bra{\mu'_u\nu'_d}.
\end{align}
Here, $\mu,\nu,\mu',\nu'=H,V$, and the parameters $I_{\mu\nu},J_{\mu\nu}^{\mu'\nu'}$ and $\phi_{\mu\nu}^{\mu'\nu'}$ are all real quantities obeying $0\leq I_{\mu\nu}\leq 1$, $0\leq J_{\mu\nu}^{\mu'\nu'}\leq 1$, and $0\leq \phi_{\mu\nu}^{\mu'\nu'}\leq 2\pi$. In order that Eq.~(\ref{q-state}) represents a valid density operator, the parameters must satisfy the relations $\phi_{\mu\nu}^{\mu'\nu'}=-\phi_{\mu'\nu'}^{\mu\nu}$, $J_{\mu\nu}^{\mu'\nu'}=J_{\mu'\nu'}^{\mu\nu}$, $J_{\mu\nu}^{\mu\nu}=1$, $\sum_{\mu,\nu}^{H,V}I_{\mu\nu}=1$, and $\phi_{\mu\nu}^{\mu\nu}=0$. One can check that $\dm$ is Hermitian, positive semidefinite and has a unit trace. 
\par
We call $I_{\mu\nu}$ an \emph{intensity parameter}, $J_{\mu\nu}^{\mu'\nu'}$ an \emph{indistinguishability parameter}, and $\phi_{\mu\nu}^{\mu'\nu'}$ a \emph{phase parameter} (see discussion in Sec.~\ref{subsec:state-para} below). It is to be noted that in the most general scenario, the density operator given by Eq.~(\ref{q-state}) has 15 free parameters. 

\subsection{State parameters from the perspective of state creation}\label{subsec:state-para}
\begin{figure*}
	\centering
	\includegraphics[width=\linewidth]{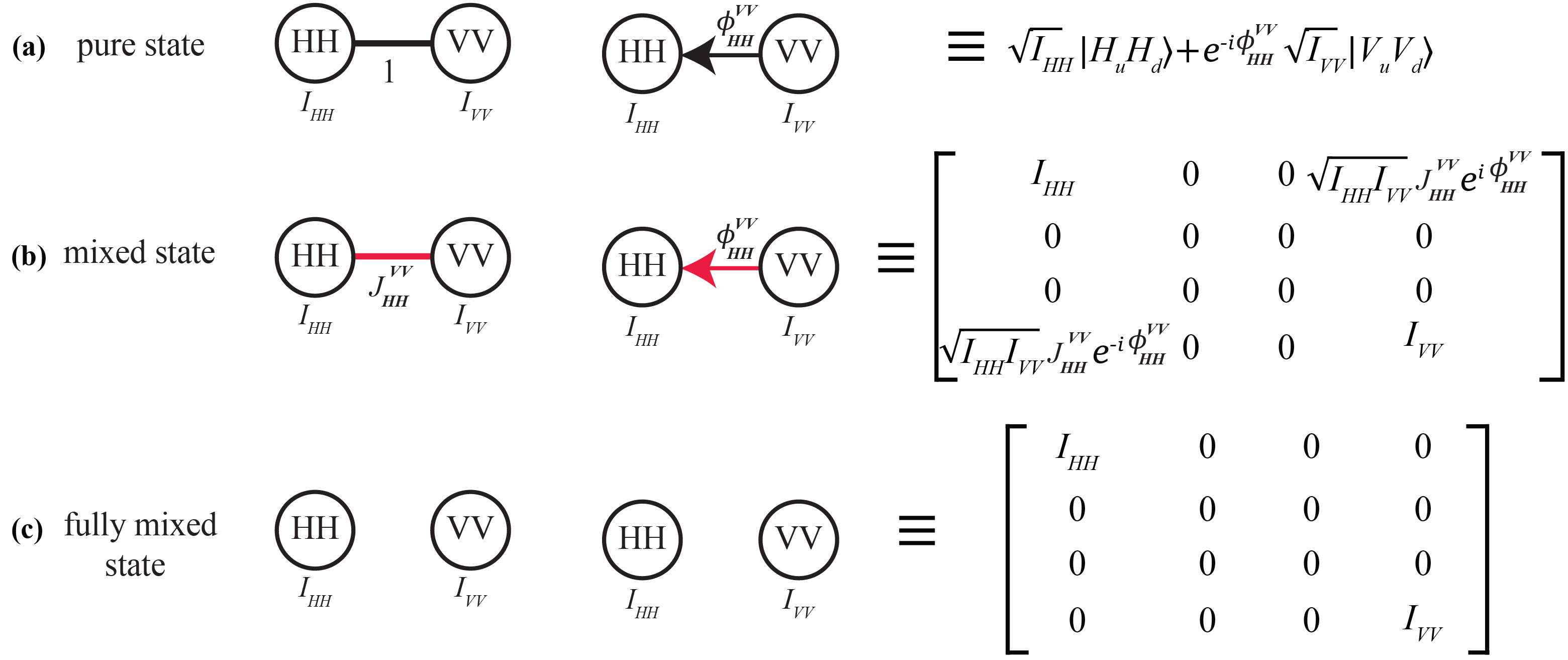}
	\caption{Graph theory representation of two-qubit states, illustrated with three states derived from a Bell state. A graph representing a state consists of two weighted subgraphs: one undirected and one directed. Vertices represent emissions, the weight of an undirected edge represents an indistinguishability parameter, and the weight of a directed edge represents a phase parameter. Colors of edges corresponding to same pair of emissions are kept same for two subgraphs belonging to the same graph. (a) Pure state. Emissions of $\ket{H_u H_d}$ and $\ket{V_u V_d}$ are mutually coherent, i.e., $J_{HH}^{VV}=1$. The color of edges are chosen black for mutually coherent emissions. If $\phi_{HH}^{VV}$ were replaced by $-\phi_{HH}^{VV}$, the direction of the directed edge would be opposite. (b) Mixed state. Emissions of $\ket{HH}$ and $\ket{VV}$ may not be fully coherent, i.e., $0\leq J_{HH}^{VV}\leq 1$. The color of the edges are not black in this case. (c) Fully mixed state. Emissions of $\ket{HH}$ and $\ket{VV}$ are mutually incoherent, i.e., $J_{HH}^{VV}=0$ and $\phi_{HH}^{VV}$ is irrelevant. This case is represented by isolated vertices (no edges).}
	\label{notation-1}
\end{figure*}
\begin{table}[b]		
	\setlength{\tabcolsep}{10pt} 
	\renewcommand{\arraystretch}{1} 
	\begin{tabular}{c | c} 
		\hline\hline 
		\textbf{Quantum state} & \textbf{Graph} \\[2 pt] 
		\hline 
		emission & vertex \\
  \hline
  fully coherent emission  & black edge \\
        ($J_{\mu\nu}^{\mu'\nu'}=1$)\\
    \hline
		partially coherent emission  & colored (not black) \\
   ($0<J_{\mu\nu}^{\mu'\nu'}<1$) &  edge \\   
  \hline
  incoherent emission  & no edge  \\
  ($J_{\mu\nu}^{\mu'\nu'}=0$) &(isolated vertices)\\
  \hline
		indistinguishability& weight of \\
         parameter & undirected edge\\
  \hline
   phase  & weight of \\
   parameter & directed edge\\
		\hline\hline
	\end{tabular}
	\caption{Rules of graph theory-based representation of a two-qubit state.}\label{pic-rep-table}
\end{table}
An understanding of the state parameters $I_{\mu\nu}$, $J_{\mu\nu}^{\mu'\nu'}$, and $\phi_{\mu\nu}^{\mu'\nu'}$ from the perspective of state creation would be useful to study the evolution of a mixed quantum state through the interferometer.
\par
Let us first consider a special case of Eq.~(\ref{q-state}), in which $I_{HH}\neq 0$, $I_{VV}\neq 0$, $\phi^{VV}_{HH}\neq 0$, $J^{VV}_{HH}=1$, and rest of the parameters are zero. This state has the form $\dm=I_{HH}\ket{H_u H_d}\bra{H_u H_d}+I_{VV}\ket{V_u V_d}\bra{V_u V_d}+\sqrt{I_{HH}I_{VV}}e^{i\phi^{VV}_{HH}}\ket{H_u H_d}\bra{V_u V_d}+\sqrt{I_{HH}I_{VV}}e^{-i\phi^{VV}_{HH}}\ket{V_u V_d}\bra{H_u H_d}$. This quantum state is pure (i.e. $\dm=\ket{\psi}\bra{\psi}$) and can be equivalently represented by
\begin{align}\label{p-state}
\ket{\psi}=\sqrt{I_{HH}}\ket{H_u H_d}+e^{-i\phi_{HH}^{VV}}\sqrt{I_{VV}}\ket{V_u V_d},
\end{align}
where $I_{HH}+I_{VV}=1$. The principle behind the creation of this state by any physical source can be stated as follows: the basic states $\ket{H_u H_d}$ and $\ket{V_u V_d}$ are \emph{coherently} emitted in such a way that the spatial modes for a given pair of photons are indistinguishable \cite{kwiat1999ultrabright,kwiat1995new}. The probability of emitting states $\ket{H_uH_d}$ and $\ket{V_uV_d}$ are proportional to $I_{HH}$ and $I_{VV}=1-I_{HH}$, respectively. We call $I_{HH}$ and $I_{VV}$ \emph{intensity parameters}. The fact that the emission of $\ket{H_u H_d}$ and $\ket{V_u V_d}$ are fully coherent to each other is quantitatively represented by $J^{VV}_{HH}=1$. Note that if $J^{VV}_{HH}$ were less than $1$, these two emissions would not have been mutually coherent. We call $J^{VV}_{HH}$ an \emph{indistinguishability parameter} since quantum indistinguishability is often connected with how coherent the two fields are. The quantity $\phi^{VV}_{HH}$ represents the phase difference between emitted $\ket{H_u H_d}$ and $\ket{V_u V_d}$ states (precisely, between corresponding fields). 
Let us call $\phi^{VV}_{HH}$ a phase parameter.
\par
The principle of generating the state given by Eq.~(\ref{p-state}) can be visualized by a graph consisting of two subgraphs (Fig.~\ref{notation-1}a): an undirected weighted sub-graph corresponding to the indistinguishability parameter, and a directed weighted sub-graph corresponding to the phase parameter. For both subgraphs, the two vertices (encircled $HH$ and $VV$) represent emissions of the two basic states $\ket{H_u H_d}$ and $\ket{V_u V_d}$. In the undirected subgraph (Fig.~\ref{notation-1}a, left), the \emph{black} edge implies that emissions of $\ket{H_u H_d}$ and $\ket{V_u V_d}$ are fully coherent to each other. The number ``1'', which represents the value of $J_{HH}^{VV}$ in this case, is the weight of this edge. In the directed subgraph (Fig.~\ref{notation-1}a, middle), the weight of the black directed edge represents the phase difference $\phi_{HH}^{VV}$ between $\ket{H_u H_d}$ and $\ket{V_u V_d}$. The direction of the edge corresponds to the sign of the phase: if $\phi_{HH}^{VV}$ were replaced by $\phi_{VV}^{HH}$, the direction of the edge would be opposite because $\phi_{VV}^{HH}=-\phi_{HH}^{VV}$. The black color of the edge once again signifies the fact that the emissions of $\ket{H_u H_d}$ and $\ket{V_u V_d}$ are mutually coherent.
\par
We now slightly generalize the state in Eq.~(\ref{p-state}) by considering the possibility that emissions of $\ket{H_u H_d}$ and $\ket{V_u V_d}$ may not be mutually coherent. In this case, the indistinguishability parameter, $J_{HH}^{VV}$, can take any value between $0$ and $1$. The quantum state is now no longer pure in general and, therefore, we must use a density operator to represent it. Without any loss of generality, the density operator can be represented in the form given by Fig.~\ref{notation-1}b, right. This density operator is a special case of Eq.~(\ref{q-state}), in which state parameters other than $I_{HH}$, $I_{VV}$ $J_{HH}^{VV}$, and $\phi_{HH}^{VV}$ are zero. We once again can use a graph made of two subgraphs to represent this state (Fig.~\ref{notation-1}b). In this case, we use edges with distinct colors for distinct values of $J_{HH}^{VV}$ (weight).  In the undirected subgraph (Fig.~\ref{notation-1}b, left), we have chosen the edge color red for the purpose of illustration. When $J_{HH}^{VV}=1$, this density operator represents the same state given by Eq.~(\ref{p-state}), i.e., a black edge is a special case of an edge of arbitrary color. 
\par
It is once again possible to identify $\phi_{HH}^{VV}$ as the phase difference between quantum fields created by the emissions of $\ket{H_uH_d}$ and $\ket{V_uV_d}$. Let us note that $\ket{H_uH_d}=\hat{a}_u^\dag(H)\hat{a}_d^\dag(H)\ket{\text{vac}}$ and $\ket{V_uV_d}=\hat{a}_u^\dag(V)\hat{a}_d^\dag(V)\ket{\text{vac}}$, where $\ket{\text{vac}}$ represents the vacuum state, $\dag$ represents Hermitian conjugation and $\hat{a}_u(\mu)$ is the annihilation operator for a single $u$-photon with polarization $\mu$, with $\mu=H,$ $V$.  It is, therefore, possible to express the density matrix element $\sqrt{I_{HH}I_{VV}}J^{VV}_{HH}e^{i\phi^{VV}_{HH}}\ket{H_uH_d}\bra{V_uV_d}$ as 
\begin{align}
    &\sqrt{I_{HH}I_{VV}}J^{VV}_{HH} \nonumber \\ &\times \left\{\hat{a}_u(H)\hat{a}_d(H)\right\}^\dag\ket{\text{vac}}\bra{\text{vac}}\big\{\hat{a}_u(V)\hat{a}_d(V)e^{i\phi_{HH}^{VV}}\big\},
\end{align}
for any non-zero value of $J_{HH}^{VV}$. We observe that $\phi_{HH}^{VV}$ appears as the phase difference between the \emph{bi-photon quantum fields} corresponding to $\ket{H_uH_d}$ and $\ket{V_uV_d}$. In the special case, when $J^{VV}_{HH}=0$, this matrix element becomes zero and consequently, $\phi_{HH}^{VV}$ does not have any physical meaning. 
\par
In the directed subgraph (Fig.~\ref{notation-1}b, middle), the edge color is kept same as that in the corresponding undirected subgraph. The weight of the directed edge represents the phase parameter $\phi_{HH}^{VV}$ and  the direction of the edge is chosen following the same convention as in the case of the pure state (Fig.~\ref{notation-1}a). 
\par
When $J^{VV}_{HH}=0$, the emissions of $\ket{H_u H_d}$ and $\ket{V_u V_d}$ are fully incoherent to each other and the quantum state becomes fully mixed [Fig.~\ref{notation-1}c, right]. No phase parameters exist in this case. Both subgraphs now contain only isolated vertices (no edge) as shown in Fig.~\ref{notation-1}c. Therefore, the case of no edge is also a special case of an edge of arbitrary color.
\begin{figure}[htbp]
	\centering
	\includegraphics[width=\linewidth]{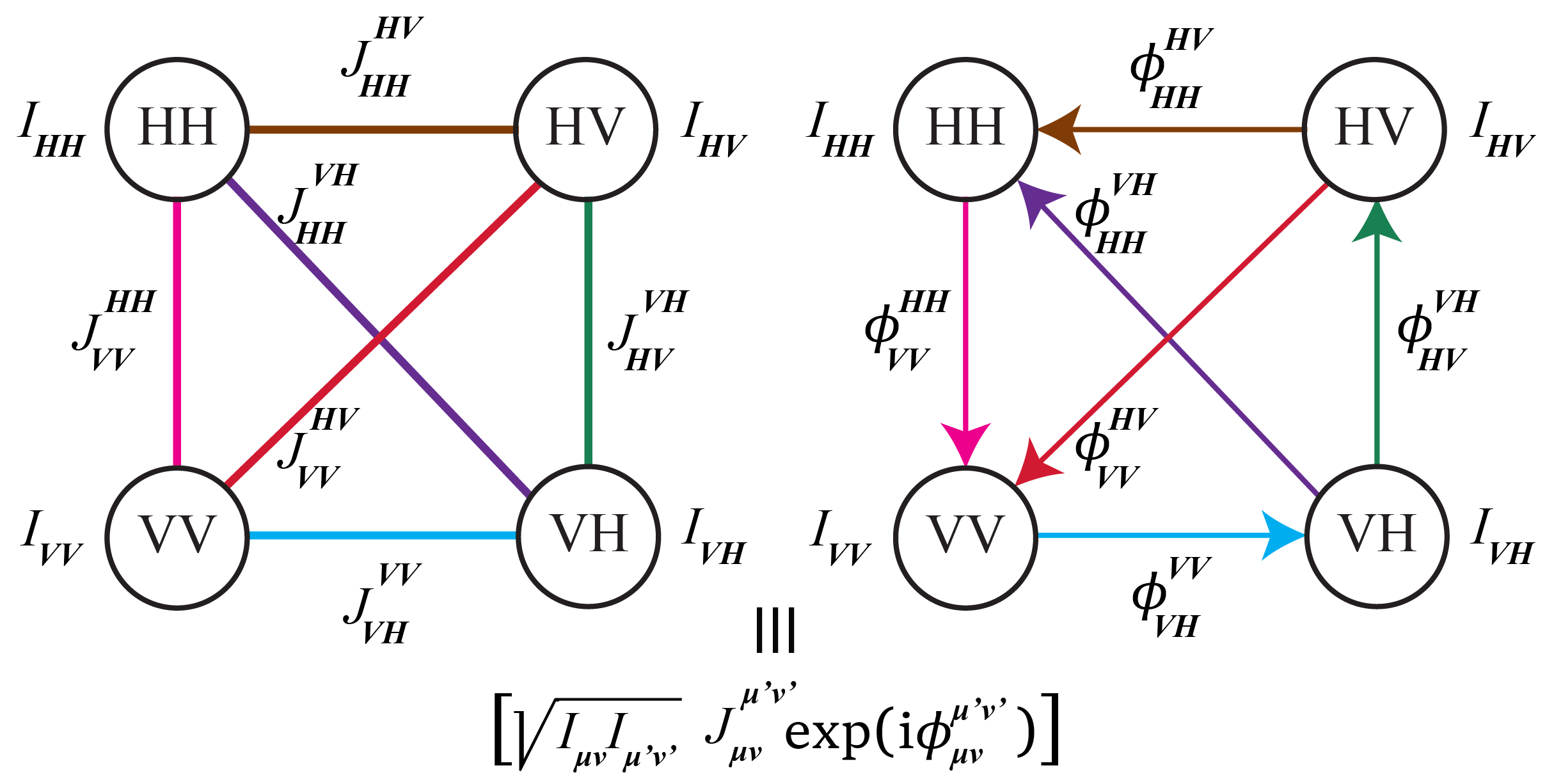}
	\caption{Graph representing an arbitrary two-qubit state with 15 free parameters. Emissions of $\ket{H_uH_d}$, $\ket{H_uV_d}$, $\ket{V_uH_d}$ and $\ket{V_uV_d}$ are represented by four vertices. Six indistinguishability and six phase parameters are represented by the weights of undirected and directed edges, respectively. Colors of edges corresponding to same pair of emissions are kept same for the two subgraphs.}
	\label{notation-2}
\end{figure}
\par
Following these basic rules (see also Table \ref{pic-rep-table}), one can create the graph corresponding to an arbitrary two-qubit mixed state that can contain up to 15 free parameters [Eq.~(\ref{q-state})]. The graph, which consists of an undirected and a directed subgraph, is shown in Fig.~\ref{notation-2}. In this case, there are four emissions corresponding to basic states $\ket{H_uH_d}$, $\ket{H_uV_d}$, $\ket{V_uH_d}$, and $\ket{V_uV_d}$. Consequently, there are four intensity parameters, $I_{\mu\nu}$ with $\mu=H,V$ and $\nu=H,V$, corresponding to probabilities of emissions. These intensity parameters obey $I_{HH}+I_{HV}+I_{VH}+I_{VV}=1$ and therefore only 3 of them are free. There are also 6 free indistinguishability parameters $J_{\mu\nu}^{\mu'\nu'}$ that characterize how coherent the emissions of $\ket{\mu_u\nu_d}$ and $\ket{\mu'_u\nu'_d}$ are; they are represented by weights of undirected edges with distinct colors. The phase parameters corresponding to the same pair of emissions are denoted by $\phi_{\mu\nu}^{\mu'\nu'}$. One can readily check that there are 6 free phase parameters, which are represented by weights of six directed edges. Colors of edges corresponding to the same pair of emissions are chosen to be the same for the two subgraphs. Graphs shown in Fig.~\ref{notation-1} are special cases of the graph shown in Fig.~\ref{notation-2}.

\subsection{Outline of the density matrix measurement scheme}\label{subsec:outline}
\begin{figure}[htbp]  
 \centering
 	\includegraphics[width=\linewidth]{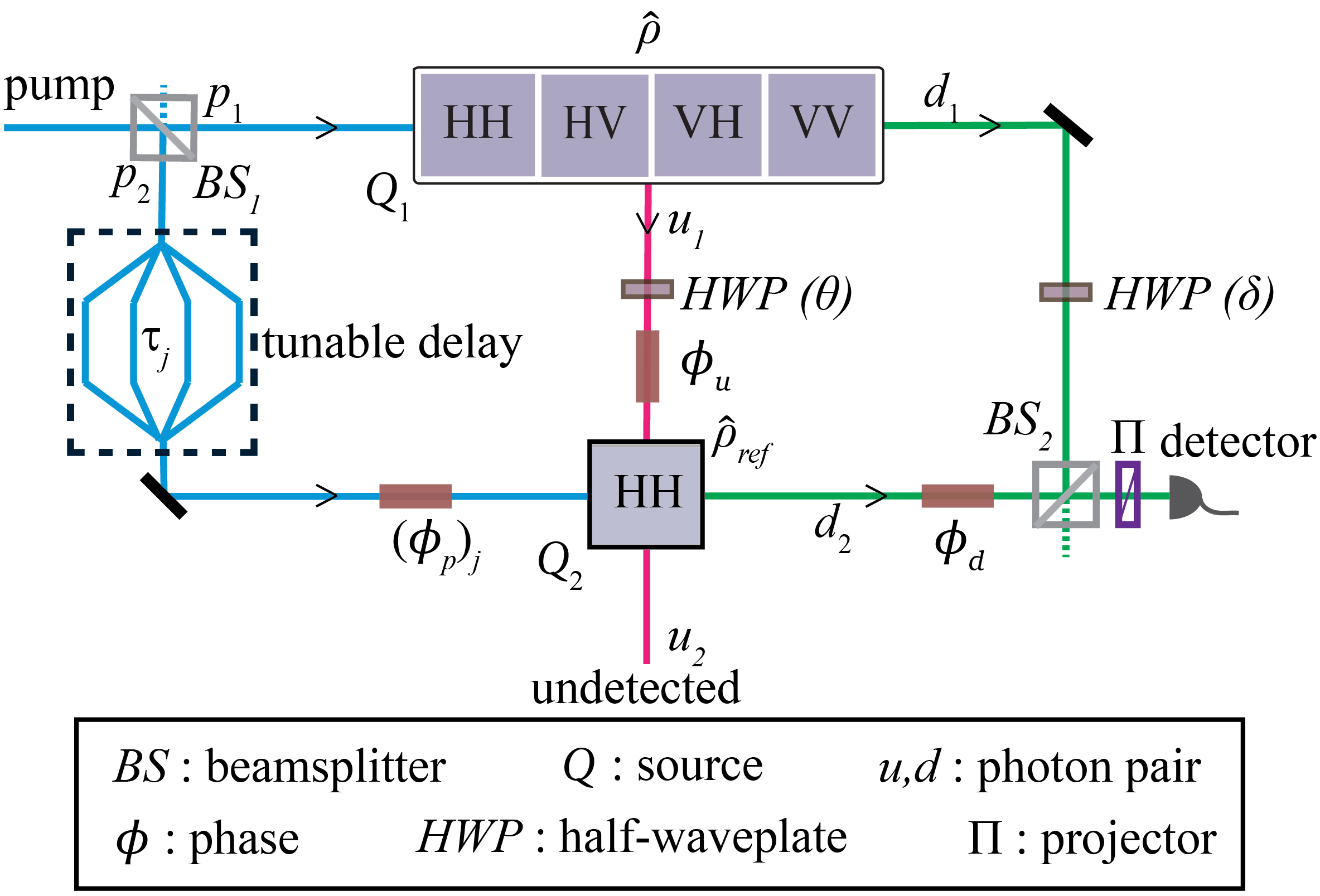}
 	\qquad
 	\caption{Two-qubit density matrix measurement (state tomography) scheme. It is an implementation of Fig.~\ref{fig:scheme}b using polarization states. To generate the most general two-qubit state ($\dm$), $Q_1$ must be able to emit $\ket{H_uH_d}$, $\ket{H_uV_d}$, $\ket{V_uH_d}$, and $\ket{V_uV_d}$, where $H$ and $V$ represent horizontal and vertical polarization. The relative probabilities of these emissions and their mutual coherence determine the form of $\dm$. Source $Q_2$ emits a known reference state $\ket{H_uH_d}$. We consider four configurations, $j=\text{A},\text{B},\text{C},\text{D}$, in which the emission from $Q_2$ is made coherent with the emission of $\ket{H_uH_d}$, $\ket{H_uV_d}$, $\ket{V_uH_d}$, and $\ket{V_uV_d}$ from $Q_1$, respectively, by introducing appropriate delays ($\tau_j$) between the corresponding pump beams $p_1$ and $p_2$. Two half-wave plates set at angles $\theta$ and $\delta$ apply unitary transformations to $u$- and $d$-photon. The $d$-photon is projected onto $H$ and $V$ polarization states before detection. Single-photon interference patterns are recorded for certain choices of $(\theta,\delta)$ in all configurations. State $\dm$ is reconstructed using these interference patterns without any postselection.} \label{fig:setup}
 \end{figure}

The details of the two-qubit density matrix measurement (quantum state tomography) scheme is illustrated by Fig.~\ref{fig:setup}. The scheme is essentially an implementation of the principle illustrated by Fig.~\ref{fig:scheme}b. Source $Q_1$ generates an arbitrary unknown two-qubit state $\dm$, which is to be reconstructed. We work with the general form of $\dm$ given by Eq.~(\ref{q-state}) such that the results are applicable to any two-qubit photonic state. Following the discussion in Sec.~\ref{subsec:state-para}, source $Q_1$ can be imagined to contain four emitters (e.g., nonlinear crystals) corresponding to the basic states $\ket{H_{u_1}H_{d_1}}$, $\ket{H_{u_1}V_{d_1}}$, $\ket{V_{u_1}H_{d_1}}$, and $\ket{V_{u_1}V_{d_1}}$, where subscript 1 refers to source $Q_1$. Therefore, 
 \begin{subequations}\label{q1-state-params}
 \begin{align}
& J_{\mu_1\nu_1}^{\mu'_1\nu'_1}=J_{\mu\nu}^{\mu'\nu'}=J^{\mu\nu}_{\mu'\nu'}, \label{q1-state-params:a} \\ &\phi_{\mu_1\nu_1}^{\mu'_1\nu'_1}=\phi_{\mu\nu}^{\mu'\nu'}=-\phi^{\mu\nu}_{\mu'\nu'} \label{q1-state-params:b}.
 \end{align}
 \end{subequations}
 We stress that the results are independent of the internal structure of the source and applicable to any source capable of generating the state $\dm$.
\par
Source $Q_2$ emits a known reference state $\dm_{ref}=\ket{\psi_{ref}}\bra{\psi_{ref}}$. We choose \cite{ref-state-note}
\begin{align}\label{q-ref-state}
\ket{\psi_{ref}}=\ket{H_{u_2}H_{d_2}},
\end{align}
where subscript 2 refers to source $Q_2$. 
\par
We consider four configurations: A, B, C, and D. In configuration A, the emission of the state $\ket{H_{u_2}H_{d_2}}$ from $Q_2$ is set fully coherent to the emission of state $\ket{H_{u_1}H_{d_1}}$ from $Q_1$. Likewise, in configurations B, C, and D, the emission of $\ket{H_{u_2}H_{d_2}}$ is made fully coherent to emissions of $\ket{H_{u_1}V_{d_1}}$, $\ket{V_{u_1}H_{d_1}}$, and $\ket{V_{u_1}V_{d_1}}$, respectively. These configurations can be realized, for example, by introducing a tunable optical delay between the two mutually coherent pump beams $p_1$ and $p_2$ (Fig.~\ref{fig:setup}). We stress that no prior information about the quantum state $\dm$ [Eq.~(\ref{q-state})] is required to identify any of the configurations. We show in Appendix \ref{app:finding-configs} that each configuration can be uniquely identified using single-photon counting rates measured by the detector.
\par
Unitary transformations $\hat{O}_1$ and $\hat{O}_2$ (Fig.~\ref{fig:scheme}b), acting on photons $u$ and $d$, respectively, are chosen as half-wave plate transformations (Fig.~\ref{fig:setup}). That is, $\hat{O}_1(\theta)=HWP(\theta)$ and $\hat{O}_2(\delta)=\textit{HWP}(\delta)$, where
\begin{align}\label{u-transform}
\textit{HWP}(\gamma)=\begin{pmatrix}
\cos 2\gamma & \sin 2\gamma \\
\sin 2\gamma & -\cos 2\gamma
\end{pmatrix}, \quad 0\leq\gamma\leq\pi.
\end{align}
Here, $\gamma=\theta,\delta$ represents half-wave plate angles. 
\par
Beams $d_1$ and $d_2$ are superposed by a beamsplitter. A $d$-photon emerging from the beamsplitter is projected onto $H$ or $V$ polarization state and then detected (Fig.~\ref{fig:setup}). We stress again that no coincidence measurement or postselection is performed; only single-photon counting rate (intensity) is measured.
We show below how to determine the entire density matrix $\dm$ from the single-photon counting data.

\subsection{Quantum state generated by emissions of the two sources}\label{subsec:total-quantum-state}
A method to study the evolution of simple two-qubit mixed states through a path identity-based interferometer was introduced in Ref.~\cite{lahiri2021characterizing} in the context of measuring entanglement in two-qubit generalized Bell states. Here, we generalize the method to an arbitrary two-qubit state. In Ref.~\cite{lahiri2021characterizing}, it was assumed that the two sources in the interferometer are identical, i.e., they generate the same quantum state. In our case, however, the two sources are not identical: $Q_1$ generates $\dm$ and $Q_2$ generates $\dm_{\text{ref}}$. The approach we develop here is readily applicable to the case in which the sources are identical.
\par
As mentioned in Sec.~\ref{sec:principle}, the probability of having quantum states occupied by more than two photons is negligible. For simplicity, we assume that the sources ($Q_1$ and $Q_2$) in Fig.~\ref{fig:setup} emit with equal probability. (More general situations are discussed in Appendix \ref{app:gen-treatment-loss}.)
In order to construct the quantum state generated by the two sources, let us note that there are in total 5 basic emissions in this case: four of these correspond to emissions of $\ket{H_{u_1}, H_{d_1}}$, $\ket{H_{u_1}, V_{d_1}}$, $\ket{V_{u_1}, H_{d_1}}$, and $\ket{V_{u_1}, V_{d_1}}$ from $Q_1$, and the fifth one corresponds to the emission of $\ket{H_{u_2}, H_{d_2}}$ from $Q_2$. By identifying intensity parameters for each emission, and indistinguishability and phase parameters for each pair of emissions, we can write down the density operator for each configuration as
\begin{align}\label{q-state-total}
&(\dmt_{du})_j=\frac{1}{2}\sum_{\mu,\nu}^{H,V} \sum_{\mu',\nu'}^{H,V} \sqrt{I_{\mu\nu}I_{\mu'\nu'}}J_{\mu\nu}^{\mu'\nu'}\exp\{i\phi_{\mu\nu}^{\mu'\nu'}\} \nonumber\\&\times\ket{\mu_{u_1}\nu_{d_1}} \bra {\mu'_{u_1}\nu'_{d_1}}+\frac{1}{2}\ket{H_{u_2}H_{d_2}}\bra{H_{u_2}H_{d_2}}\nonumber\\&+\frac{1}{2}\Big\{e^{i\phi_0}\sum_{\mu,\nu}^{H,V} \sqrt{I_{\mu\nu}}(J_{\mu_1\nu_1}^{H_2H_2})_j\exp[{i(\phi_{\mu_1\nu_1}^{H_2H_2}})_j]\nonumber\\&\times\ket{\mu_{u_1}\nu_{d_1}} \bra{H_{u_2}H_{d_2}}+H.c.\Big\},
\end{align}
where $j=\text{A},\text{B},\text{C},\text{D}$ represents a configuration (see Sec.~\ref{subsec:outline}), $\phi_0$ represents a phase, $H.c.$ denotes Hermitian conjugation, and we have assumed that both sources emit with equal probability. 
The first term on the right-hand side of Eq.~(\ref{q-state-total}) is solely due to emissions from $Q_1$. The first term contains indistinguishability and phase parameters of the unknown state $\dm$. These parameters do not depend on the choice of configuration. The second term is due to the emission of the known state $\ket{H_{u_2}, H_{d_2}}$. The second term also does not depend on the choice of configuration. However, the third term and its Hermitian conjugate may change from one configuration to another. This is because it contains indistinguishability and phase parameters corresponding to emissions from distinct sources ($Q_1$ and $Q_2$). We now show how to express these parameters in terms of the indistinguishability and phase parameters of $\dm$.
\par
Let us first consider configuration A. In this configuration, emission of $\ket{H_{u_2}H_{d_2}}$ from $Q_2$ is fully coherent to the emission of $\ket{H_{u_1}H_{d_1}}$ from $Q_1$, i.e., 
\begin{align}\label{J-confA}
(J_{H_1 H_1}^{H_2H_2})_{\text{A}}=1.
\end{align} 
Consequently, the amount of coherence between emissions of $\ket{H_{u_2}, H_{d_2}}$ and any other basic state $\ket{\mu_{u_1}, \nu_{d_1}}$ must be equal to that of $\ket{H_{u_1}, H_{d_1}}$ and $\ket{\mu_{u_1}, \nu_{d_1}}$. That is, we have 
\begin{align}\label{J-A}
(J_{\mu_1 \nu_1}^{H_2H_2})_{\text{A}}=J^{H_1 H_1}_{\mu_1\nu_1}.
\end{align} 
Using Eqs.~(\ref{q1-state-params:a}) and (\ref{J-A}), we now readily find that 
\begin{align}\label{J-2}
(J_{\mu_1 \nu_1}^{H_2H_2})_{\text{A}}=J^{HH}_{\mu\nu}=J_{HH}^{\mu\nu}.
\end{align} 
Note that Eq.~(\ref{J-confA}) is a special case of Eq.~(\ref{J-2}) since $J^{HH}_{HH}=1$.
Considering all possible combinations of $\mu$ and $\nu$, we obtain the following relations in addition to Eq.~(\ref{J-confA}):
\begin{subequations}\label{J-3}
\begin{align}
&(J_{H_1 V_1}^{H_2H_2})_{\text{A}}=J_{HH}^{HV}, \label{J-3:a} \\
&(J_{V_1 H_1}^{H_2H_2})_{\text{A}}=J_{HH}^{VH}, \label{J-3:b} \\
& (J_{V_1 V_1}^{H_2H_2})_{\text{A}}=J_{HH}^{VV}. \label{J-3:c}
\end{align} 
\end{subequations}
\par
\begin{figure}[htbp] \centering
    \includegraphics[width=\linewidth]{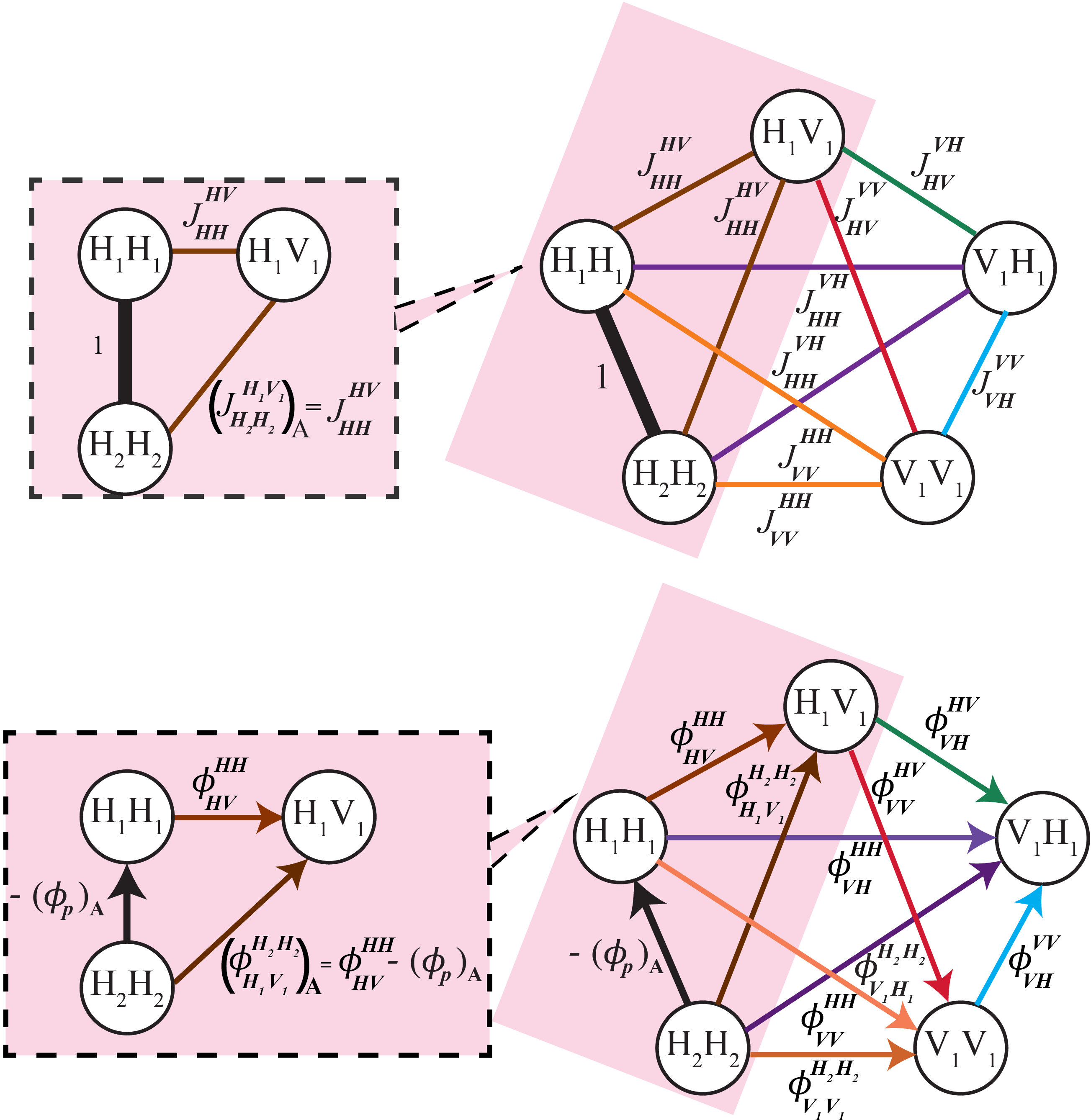}
    \qquad
    \caption{Graph representing the quantum state in configuration A before path identity is employed. (Intensity parameters are not shown.) The subgraph on the top corresponds to indistinguishability parameters [Eqs.~(\ref{J-3:a})-(\ref{J-3:c})]. The black edge connecting vertices $\text{H}_1\text{H}_1$ and $\text{H}_2\text{H}_2$ signifies that emissions of $\ket{H_{u_1}H_{d_1}}$ and $\ket{H_{u_2}H_{d_2}}$ are mutually coherent in configuration A. Consequently, edges connecting ($\text{H}_1\text{H}_1$, $\text{H}_1\text{V}_1$) and ($\text{H}_2\text{H}_2$, $\text{H}_1\text{V}_1$) have the same weight and color (see inset), which implies $(J_{H_1 V_1}^{H_2H_2})_{\text{A}}=J_{HH}^{HV}$ [Eq.~(\ref{J-3:a})]. Likewise, Eqs.~(\ref{J-3:b}) and (\ref{J-3:c}) can be obtained using this subgraph. The subgraph at the bottom corresponds to phase parameters [Eqs.~(\ref{phase-A-params-a})-(\ref{phase-A-params-c})]. The inset shows that the phase parameter corresponding to ($\text{H}_2\text{H}_2$, $\text{H}_1\text{V}_1$) is obtained by adding those corresponding to ($\text{H}_2\text{H}_2$, $\text{H}_1\text{H}_1$) and ($\text{H}_1\text{H}_1$, $\text{H}_1\text{V}_1$) according to the triangle law of vector addition, i.e., $(\phi_{H_1V_1}^{H_2H_2})_{\text{A}}=\phi_{HV}^{HH}-(\phi_p)_{\text{A}}$ [Eq.~(\ref{phase-A-params-a})]. Likewise, Eqs.~(\ref{phase-A-params-b}) and (\ref{phase-A-params-c}) can be obtained using this subgraph.}.  \label{config-a-graph}
\end{figure}

The process of obtaining Eqs.~(\ref{J-3:a})-(\ref{J-3:c}) can be visualized using the subgraph given in Fig.~\ref{config-a-graph} (top). Equation (\ref{J-confA}) is represented by the black edge with weight $1$ connecting vertices $\text{H}_1\text{H}_1$ and $\text{H}_2\text{H}_2$. This edge is made bold for the ease of identification.
An undirected edge joining vertices $\text{H}_2\text{H}_2$ and $\mu_1\nu_1$ $\neq$ $\text{H}_1\text{H}_1$ has the same weight and color as the edge joining $\text{H}_1\text{H}_1$ and $\mu_1\nu_1$. The process of obtaining Eq.~(\ref{J-3:a}) is illustrated in the inset.
\par
We now consider the phase parameters in configuration A. We recall from Sec.~\ref{subsec:state-para} that a phase parameter $\phi_{\mu\nu}^{\mu'\nu'}$ represents the phase difference between the bi-photon fields corresponding to $\ket{\mu_u\nu_d}$ and $\ket{\mu'_u\nu'_d}$. In this notation, $(\phi_{H_1H_1}^{H_2H_2})_{\text{A}}$ represents the phase difference between fields corresponding to states $\ket{H_{u_1}H_{d_1}}$ and $\ket{H_{u_2}H_{d_2}}$ in configuration A. This phase difference must be given by the phase difference $(\phi_p)_{\text{A}}$ between pump fields at the corresponding emitters. We thus have in configuration A,
\begin{align}\label{p-phase-A}
    (\phi_{H_1H_1}^{H_2H_2})_{\text{A}}=-(\phi_{H_2H_2}^{H_1H_1})_{\text{A}}=-(\phi_p)_{\text{A}}.
\end{align}
We note that $(\phi_p)_{\text{A}}$ can be tuned in an experiment. 
\par
The phase difference $(\phi_{\mu_1\nu_1}^{H_2H_2})_{\text{A}}$ between the fields corresponding to $\ket{\mu_{u_1}\nu_{d_1}}$ and $\ket{H_{u_2}H_{d_2}}$ can now be determined using the standard rule of addition of phases in an optical interferometer. Since the phase difference between $\ket{\mu_{u_1}\nu_{d_1}}$ and $\ket{H_{u_1}H_{d_1}}$ is $\phi_{\mu\nu}^{HH}$ [see Eq.~(\ref{q1-state-params:b})], using Eq.~(\ref{p-phase-A}), we find that the phase difference between $\ket{\mu_{u_1}\nu_{d_1}}$ and $\ket{H_{u_2}H_{d_2}}$ in configuration A is given by
\begin{align}\label{phase-param-1-A}
(\phi_{\mu_1\nu_1}^{H_2H_2})_{\text{A}}=\phi_{\mu\nu}^{HH}-(\phi_p)_{\text{A}}=-\phi^{\mu\nu}_{HH}-(\phi_p)_{\text{A}}.
\end{align}
We observe that Eq.~(\ref{p-phase-A}) is a special case of Eq.~(\ref{phase-param-1-A}) since $\phi_{HH}^{HH}=0$. Using Eq.~(\ref{phase-param-1-A}) and considering all the possible combinations of $\mu$ and $\nu$,  we obtain the following relations in addition to Eq.~(\ref{p-phase-A}) in configuration A:
\begin{subequations}\label{phase-params-A}
    \begin{align}
        &(\phi_{H_1V_1}^{H_2H_2})_{\text{A}}=\phi_{HV}^{HH}-(\phi_p)_{\text{A}}, \label{phase-A-params-a}\\
        &(\phi_{V_1H_1}^{H_2H_2})_{\text{A}}=\phi_{VH}^{HH}-(\phi_p)_{\text{A}}, \label{phase-A-params-b}\\
        &(\phi_{V_1V_1}^{H_2H_2})_{\text{A}}=\phi_{VV}^{HH}-(\phi_p)_{\text{A}}.\label{phase-A-params-c}
    \end{align}
\end{subequations}
\par
Equations (\ref{phase-A-params-a})-(\ref{phase-A-params-c}) are represented by the directed subgraph in Fig.~\ref{config-a-graph} (bottom). Let us consider the highlighted part of the graph (inset); if the weights of the directed edges are added according to the triangle law of vector addition, we obtain Eq.~(\ref{phase-A-params-a}). Equations (\ref{phase-A-params-b}) and (\ref{phase-A-params-c}) can also be obtained in similar manner as evident from this graph.
\par
Following the same procedure all indistinguishability parameters and phase parameters corresponding to emissions from distinct sources ($Q_1$ and $Q_2$) in configurations B, C, and D are expressed in terms of the indistinguishability and phase parameters of $\dm$. They are given in Table \ref{tab:1} and the corresponding graphs are in Appendix \ref{app:q-state-full}. We emphasize that the relationships involving indistinguishability and phase parameters (Table~\ref{tab:1}) remain unchanged even when sources $Q_1$ and $Q_2$ emit with unequal probabilities. This is because these parameters are independent of the emission probabilities of the sources.
	\par
	Using Eq.~(\ref{q-state-total}) and the relations given by Table \ref{tab:1}, one can determine the quantum state jointly produced by $Q_1$ and $Q_2$ in any configuration. For example, the quantum state in configuration A is given by
	\begin{align}\label{q-state-total-A}
	&(\dmt_{du})_{\text{A}}=\frac{1}{2}\Big\{\sum_{\mu,\nu}^{H,V} \sum_{\mu',\nu'}^{H,V} \sqrt{I_{\mu\nu}I_{\mu'\nu'}}J_{\mu\nu}^{\mu'\nu'}\exp\{i\phi_{\mu\nu}^{\mu'\nu'}\}  \nonumber\\&\times\ket{\mu_{u_1}\nu_{d_1}}\bra {\mu'_{u_1}\nu'_{d_1}}+\ket{H_{u_2}H_{d_2}}\bra{H_{u_2}H_{d_2}}\nonumber\\&+\Big[e^{i\phi_0}\sum_{\mu,\nu}^{H,V} \sqrt{I_{\mu\nu}}J_{\mu\nu}^{HH}\exp\{i(\phi_{\mu\nu}^{HH}-(\phi_p)_{\text{A}})\} \nonumber\\&\times\ket{\mu_{u_1}\nu_{d_1}} \bra{H_{u_2}H_{d_2}}+H.c.\Big]\Big\}.
	\end{align}
	\par
	We have thus described a method to determine the density operator representing a photon pair generated by combined emissions of $Q_1$ and $Q_2$. Note that the alignment of $u$-photon beams (path identity) is not yet considered.
	\begin{table*}[t]		
		\setlength{\tabcolsep}{7.5 pt} 
		\renewcommand{\arraystretch}{1} 
		\begin{tabular}{l  l  l  l} 
			\hline\hline 
			Configuration A & Configuration B  &  Configuration C  & Configuration D\\[2 pt] 
			\hline 
			$(J_{H_1H_1}^{H_2H_2})_{\text{A}}=1$ & $(J_{H_1H_1}^{H_2H_2})_{\text{B}}=J_{HH}^{HV}$ & $(J_{H_1H_1}^{H_2H_2})_{\text{C}}=J_{HH}^{VH}$ & $(J_{H_1H_1}^{H_2H_2})_{\text{D}}=J_{HH}^{VV}$\\$(\phi_{H_1H_1}^{H_2H_2})_{\text{A}}=-(\phi_p)_{\text{A}}$ & $(\phi_{H_1H_1}^{H_2H_2})_{\text{B}}=\phi_{HH}^{HV} -(\phi_p)_{\text{B}}$& $(\phi_{H_1H_1}^{H_2H_2})_{\text{C}}=\phi_{HH}^{VH}-(\phi_p)_{\text{C}}$ & $(\phi_{H_1H_1}^{H_2H_2})_{\text{D}}=\phi_{HH}^{VV}-(\phi_p)_{\text{D}}$\\[1ex] 
            \hline
			$(J_{H_1V_1}^{H_2H_2})_{\text{A}}=J_{HV}^{HH}$ & $(J_{H_1V_1}^{H_2H_2})_{\text{B}}=1$ & $(J_{H_1V_1}^{H_2H_2})_{\text{C}}=J_{HV}^{VH}$ & $(J_{H_1V_1}^{H_2H_2})_{\text{D}}=J_{HV}^{VV}$\\$(\phi_{H_1V_1}^{H_2H_2})_{\text{A}}=\phi_{HV}^{HH}-(\phi_p)_{\text{A}}$ & $(\phi_{H_1V_1}^{H_2H_2})_{\text{B}}=-(\phi_p)_{\text{B}}$ & $(\phi_{H_1V_1}^{H_2H_2})_{\text{C}}=\phi_{HV}^{VH}-(\phi_p)_{\text{C}}$ & $(\phi_{H_1V_1}^{H_2H_2})_{\text{D}}=\phi_{HV}^{VV}-(\phi_p)_{\text{D}}$\\[1ex] 
			\hline
			$(J_{V_1H_1}^{H_2H_2})_{\text{A}}=J_{VH}^{HH}$ & $(J_{V_1H_1}^{H_2H_2})_{\text{B}}=J_{VH}^{HV}$ & $(J_{V_1H_1}^{H_2H_2})_{\text{C}}=1$ & $(J_{V_1H_1}^{H_2H_2})_{\text{D}}=J_{VH}^{VV}$\\$(\phi_{V_1H_1}^{H_2H_2})_{\text{A}}=\phi_{VH}^{HH}-(\phi_p)_{\text{A}}$ & $(\phi_{V_1H_1}^{H_2H_2})_{\text{B}}=\phi_{VH}^{HV}-(\phi_p)_{\text{B}}$ & $(\phi_{V_1H_1}^{H_2H_2})_{\text{C}}=-(\phi_p)_{\text{C}}$ & $(\phi_{V_1H_1}^{H_2H_2})_{\text{D}}=\phi_{VH}^{VV}-(\phi_p)_{\text{D}}$\\[1ex] 
			\hline
			$(J_{V_1V_1}^{H_2H_2})_{\text{A}}=J_{VV}^{HH}$ & $(J_{V_1V_1}^{H_2H_2})_{\text{B}}=J_{VV}^{HV}$ & $(J_{V_1V_1}^{H_2H_2})_{\text{C}}=J_{VV}^{VH}$ & $(J_{V_1V_1}^{H_2H_2})_{\text{D}}=1$\\
            $(\phi_{V_1V_1}^{H_2H_2})_{\text{A}}=\phi_{VV}^{HH}-(\phi_p)_{\text{A}}$ & $(\phi_{V_1V_1}^{H_2H_2})_{\text{B}}=\phi_{VV}^{HV}-(\phi_p)_{\text{B}}$ & $(\phi_{V_1V_1}^{H_2H_2})_{\text{C}}=\phi_{VV}^{VH}-(\phi_p)_{\text{C}}$ & $(\phi_{V_1V_1}^{H_2H_2})_{\text{D}}=-(\phi_p)_{\text{D}}$\\[1ex] 
			\hline\hline
		\end{tabular}
		\caption{Indistinguishability and Phase parameters in all four configurations.}\label{tab:1}
	\end{table*}
\subsection{Path identity and unitary transformation}\label{subsec:alignment-conditions}
 As mentioned in Sec.~\ref{sec:principle}, the paths of $u$-photon emerging from the two sources are made identical by sending beam $u_1$ through $Q_2$ and overlapping beams $u_1$ and $u_2$ (Fig.~\ref{fig:setup}). Such an alignment is called path identity \cite{hochrainer2022quantum}. A $u$-photon in beam $u_1$ undergoes a unitary transformation \textit{HWP}$(\theta)$, where \textit{HWP} is defined by Eq.~(\ref{u-transform}). Consequently, the quantum fields corresponding to the $u$-photon at $Q_1$ and $Q_2$ become related by
	\begin{align}\label{algn-operats}
	&\hat{a}_{u_2}(H)=e^{i\phi_u}\Big[\cos (2\theta)~\hat{a}_{u_1}(H)+\sin (2\theta)~\hat{a}_{u_1}(V)\Big],
	\end{align}
	where $\hat{a}$ represents the annihilation operator and $\phi_u$ is the phase change due to propagation from $Q_1$ to $Q_2$ along beam $u_1$. Since $Q_2$ does not generate $V$ component of the field, the operator $\hat{a}_{u_2}(V)$ is not relevant. 
	Let us note that $\hat{a}^\dag_{u_j}(\mu)\ket{\text{vac}}=\ket{\mu_{u_j}}$, where $\ket{\text{vac}}$ refers to the vacuum state and $\hat{a}^\dag$ represent the photon creation operator. Therefore, Eq.~(\ref{algn-operats}) can also be expressed as
	\begin{align}\label{algn}
	&\ket{H_{u_2}}=e^{-i\phi_u}\left[\cos (2\theta)~ \ket{H_{u_1}}+\sin (2\theta)~ \ket{V_{u_1}}\right].
	\end{align}
	Equations (\ref{algn-operats}) and (\ref{algn}) are valid for all configurations. 
\par
Applying the relationship between kets given by Eq.~(\ref{algn}) to the states generated by coherent emission of the two sources [e.g. Eq.~(\ref{q-state-total-A})], we obtain expressions for the density operator in all configurations. These two-photon density operators can be used to determine the probability of detecting a $d$-photon at the detector. Alternatively, one can also use the reduced density matrices representing the $d$-photon, which are obtained by taking the partial trace of the two-photon density matrices over the undetected photon ($u$) subspace. We follow the latter approach here. It can be shown that in a given configuration such a reduced density matrix takes the form (see Appendix~\ref{app:reduced-density-operator}):
\begin{widetext}
\begin{align}\label{detected-dm}
(\dm_d)_j&=\frac{1}{2}\Big\{\sum_{\mu,\nu}^{H,V}I_{\mu\nu}\ket{\nu_{d_1}}\bra{\nu_{d_1}}+\Big[\big(\sqrt{I_{HH}I_{HV}}J_{HH}^{HV}e^{i\phi_{HH}^{HV}}+\sqrt{I_{VH}I_{VV}}J_{VH}^{VV}e^{i\phi_{VH}^{VV}}\big)\ket{H_{d_1}}\bra{V_{d_1}}+H.c.\Big]+\ket{H_{d_2}}\bra{H_{d_2}}\nonumber\\&+\Big[e^{i(\phi_0+\phi_u)}\Big(\cos(2\theta)\big(\sqrt{I_{HH}}(J_{H_1H_1}^{H_2H_2})_j  e^{i(\phi_{H_1H_1}^{H_2H_2})_j}\ket{H_{d_1}}\bra{H_{d_2}}+\sqrt{I_{HV}}(J_{H_1V_1}^{H_2H_2})_j e^{i(\phi_{H_1V_1}^{H_2H_2})_j}\ket{V_{d_1}} \bra{H_{d_2}}\big)\nonumber\\&+\sin(2\theta)\big(\sqrt{I_{VH}}(J_{V_1H_1}^{H_2H_2})_je^{i(\phi_{V_1H_1}^{H_2H_2})_j}\ket{H_{d_1}}\bra{H_{d_2}}+\sqrt{I_{VV}}(J_{V_1V_1}^{H_2H_2})_j e^{i(\phi_{V_1V_1}^{H_2H_2})_j}\ket{V_{d_1}}\bra{H_{d_2}}\big)\Big)+H.c.\Big]\Big\},
\end{align}
\end{widetext}
where $j=\text{A},\text{B},\text{C},\text{D}$ represents a configuration and the expressions for $(J_{\mu_1\nu_1}^{H_2H_2})_j$ and $(\phi_{\mu_1\nu_1}^{H_2H_2})_j$ are given in Table~\ref{tab:1}.

\subsection{Measurement}\label{subsec:phtn-detect-probty}
We now determine single-photon counting rates at the detector for each configuration. These rates are linearly proportional to probabilities of detecting a $d$-photon.
\par
As mentioned in Sec.~\ref{sec:analytical-descrpn}, beams $d_1$ and $d_2$ are superposed by a beamsplitter and one of the outputs of the beamsplitter is sent to a detector (Fig.~\ref{fig:setup}). 
A $d$-photon in beam $d_2$ undergoes a half-wave plate transformation, $HWP(\delta)$, before arriving at the beamsplitter. The form of $HWP$ is given by Eq.~(\ref{u-transform}).
The $d$-photon is projected onto $H$ or $V$ polarization state before detection. 
Therefore, the positive frequency part of the quantized electric field components at the detector are given by
\begin{align}
	&E^{(+)}_{H_d}=\frac{1}{\sqrt{2}}\big\{\hat{a}_{{d_1}}(H) \nonumber \\ &+ie^{i\phi_d}\big[\cos (2\delta) \, \hat{a}_{{d_2}}(H)+\sin (2\delta) \, \hat{a}_{{d_2}}(V) \big]\big\},\label{E-H}
    \end{align}
and
\begin{align}
	&E^{(+)}_{V_d}=\frac{1}{\sqrt{2}}\big\{\hat{a}_{{d_1}}(V) \nonumber \\ &+ie^{i\phi_d}\big[\sin (2\delta) \, \hat{a}_{{d_2}}(H)-\cos (2\delta) \, \hat{a}_{{d_2}}(V) \big]\big\},\label{E-V}
	\end{align}
where $\hat{a}_{{d_j}}(\mu)$ is the annihilation operator corresponding to a $d$-photon with polarization $\mu$ emitted from source $Q_j$, the parameter $\delta$ represents the half-wave plate angle, and $\phi_d$ is the phase difference corresponding to the difference between optical paths along beams $d_1$ and $d_2$. In any configuration, the probability of detecting a $d$-photon with polarization $\mu$ can be found by using the standard formula \cite{glauber1963quantum}
\begin{align}\label{ph-countg-rt}
[P_\mu]_j=\text{Tr}\left\{(\dm_d)_jE^{(-)}_{\mu_d}E^{(+)}_{\mu_d}\right\},
\end{align}
where $E^{(-)}_{\mu_d}=\left(E^{(+)}_{\mu_d}\right)^{\dag}$, and $j=\text{A},\text{B},\text{C},\text{D}$ refers to a configuration.
\par
Using Eqs.~(\ref{detected-dm}), (\ref{E-H}) and (\ref{ph-countg-rt}), we find that the detection probability of an $H$-polarized $d$-photon is given by
\begin{align}\label{P-H}
&[P_H(\theta,\delta)]_j=\frac{1}{4}(I_{HH}+I_{VH}) +\frac{1}{4} \cos^2 (2\delta)\nonumber \\ &+ \frac{1}{2} \cos (2\delta) \Big\{\sqrt{I_{HH}}(J_{H_1H_1}^{H_2H_2})_j \cos (2\theta) \sin\left[\phi+(\phi_{H_1H_1}^{H_2H_2})_j\right] \nonumber\\&+\sqrt{I_{VH}} (J_{V_1H_1}^{H_2H_2})_j \sin (2\theta) \sin\left[\phi+(\phi_{V_1H_1}^{H_2H_2})_j\right] \Big\},
\end{align}
where $(J_{\mu_1\nu_1}^{H_2H_2})_j$ and $(\phi_{\mu_1\nu_1}^{H_2H_2})_j$ are given by Table~\ref{tab:1} and $\phi=\phi_0+\phi_u-\phi_d$. In an experiment, $\phi$ can be varied, for example, by varying $\phi_d$. It is evident from Eq.~(\ref{P-H}) that as $\phi$ varies $P_H(\theta,\delta)$ varies sinusoidally, i.e., $P_H(\theta,\delta)$ represents a single-photon interference pattern.
\par
Similarly, using Eqs.~(\ref{detected-dm}), (\ref{E-V}) and (\ref{ph-countg-rt}), we find that the probability of detecting a $V$-polarized $d$-photon in an arbitrary configuration is given by
\begin{align}\label{P-V}
&[P_V(\theta,\delta)]_j=\frac{1}{4}(I_{HV}+I_{VV})+\frac{1}{4} \sin^2 (2\delta) \nonumber \\ &+ \frac{1}{2}\sin (2\delta) \Big\{\sqrt{I_{HV}}(J_{H_1V_1}^{H_2H_2})_j \cos (2\theta) \sin\left[\phi+(\phi_{H_1V_1}^{H_2H_2})_j\right] \nonumber\\&+\sqrt{I_{VV}} (J_{V_1V_1}^{H_2H_2})_j \sin (2\theta) \sin\left[\phi+(\phi_{V_1V_1}^{H_2H_2})_j\right]\Big\}.
\end{align}
Like $[P_H(\theta,\delta)]_j$, the probability $[P_V(\theta,\delta)]_j$ also represents a single-photon interference pattern.
\par
We observe that the expressions for single-photon detection probabilities (counting rates) given by Eqs.~(\ref{P-H}) and (\ref{P-V}) contain the unitary-transformation parameters $\theta$ and $\delta$, as well as the parameters of the unknown quantum state $\dm$. In the Sec.~\ref{subsec:state-reconstruction} below, we show how to reconstruct an arbitrary two-qubit state using these detection probabilities.
\subsection{Reconstruction of an arbitrary two-qubit density matrix}\label{subsec:state-reconstruction}
The probability of detecting a $d$-photon of a given polarization can be determined in all configurations using Eqs.~(\ref{P-H}), (\ref{P-V}), and Table~\ref{tab:1}. We choose the following combinations for the half-wave plate angles $(\theta,\delta)$: $(0,0)$, $(0,\pi/4)$, $(\pi/4,0)$, and $(\pi/4,\pi/4)$. We find that the resulting single-photon detection probabilities are given by
\begin{widetext}
\begin{subequations}
	\begin{align}
	\text{configuration A:} \quad &[P_H(0,0)]_{\text{A}}=\frac{1}{4}(I_{HH}+I_{VH}+1)+\frac{1}{2}\sqrt{I_{HH}}\sin(\phi-(\phi_p)_{\text{A}})\label{p-h-1-phase-1}, \\ 
	&[P_H(\pi/4,0)]_{\text{A}}=\frac{1}{4}(I_{HH}+I_{VH}+1)+\frac{1}{2}\sqrt{I_{VH}}J_{VH}^{HH}\sin(\phi-(\phi_p)_{\text{A}}+\phi_{VH}^{HH})\label{p-h-1-phase-2},\\
	&[P_V(0,\pi/4)]_{\text{A}}=\frac{1}{4}(I_{HV}+I_{VV}+1)+\frac{1}{2}\sqrt{I_{HV}}J_{HV}^{HH}\sin(\phi-(\phi_p)_{\text{A}}+\phi_{HV}^{HH})\label{p-v-1-phase-1},\\ 
	&[P_V(\pi/4,\pi/4)]_{\text{A}}=\frac{1}{4}(I_{HV}+I_{VV}+1)+\frac{1}{2}\sqrt{I_{VV}}J_{VV}^{HH} \sin(\phi-(\phi_p)_{\text{A}}+\phi_{VV}^{HH})\label{p-v-1-phase-2},\\[5 pt]
    \text{configuration B:} \quad &[P_H(0,0)]_{\text{B}}=\frac{1}{4}(I_{HH}+I_{VH}+1)+\frac{1}{2}\sqrt{I_{HH}}J_{HH}^{HV}\sin(\phi-(\phi_p)_{\text{B}}+\phi^{HV}_{HH})\label{p-h-2-phase-1},\\
	&[P_H(\pi/4,0)]_{\text{B}}=\frac{1}{4}(I_{HH}+I_{VH}+1)+\frac{1}{2}\sqrt{I_{VH}}J_{VH}^{HV}\sin(\phi-(\phi_p)_{\text{B}}+\phi_{VH}^{HV})\label{p-h-2-phase-2},\\
	&[P_V(0,\pi/4)]_{\text{B}}=\frac{1}{4}(I_{HV}+I_{VV}+1)+\frac{1}{2}\sqrt{I_{HV}}\sin(\phi-(\phi_p)_{\text{B}})\label{p-v-2-phase-1},\\
	&[P_V(\pi/4,\pi/4)]_{\text{B}}=\frac{1}{4}(I_{HV}+I_{VV}+1)+\frac{1}{2}\sqrt{I_{VV}}J_{VV}^{HV}\sin(\phi-(\phi_p)_{\text{B}}+\phi_{VV}^{HV})\label{p-v-2-phase-2},\\[5 pt]
    \text{configuration C:} \quad &[P_H(0,0)]_{\text{C}}=\frac{1}{4}(I_{HH}+I_{VH}+1)+\frac{1}{2}\sqrt{I_{HH}}J_{HH}^{VH}\sin(\phi-(\phi_p)_{\text{C}}+\phi^{VH}_{HH})\label{p-h-3-phase-1},\\
    &[P_H(\pi/4,0)]_{\text{C}}=\frac{1}{4}(I_{HH}+I_{VH}+1)+\frac{1}{2}\sqrt{I_{VH}}\sin(\phi-(\phi_p)_{\text{C}})\label{p-h-3-phase-2},\displaybreak\\
    &[P_V(0,\pi/4)]_{\text{C}}=\frac{1}{4}(I_{HV}+I_{VV}+1)+\frac{1}{2}\sqrt{I_{HV}}J_{HV}^{VH}\sin(\phi-(\phi_p)_{\text{C}}+\phi_{HV}^{VH})\label{p-v-3-phase-1},\\
    &[P_V(\pi/4,\pi/4)]_{\text{C}}=\frac{1}{4}(I_{HV}+I_{VV}+1)+\frac{1}{2}\sqrt{I_{VV}}J_{VV}^{VH}\sin(\phi-(\phi_p)_{\text{C}}+\phi_{VV}^{VH})\label{p-v-3-phase-2},\\[5 pt]
    \text{configuration D:} \quad & [P_H(0,0)]_{\text{D}}=\frac{1}{4}(I_{HH}+I_{VH}+1)+\frac{1}{2}\sqrt{I_{HH}}J_{HH}^{VV}\sin(\phi-(\phi_p)_{\text{D}}+\phi^{VV}_{HH})\label{p-h-4-phase-1},\\
    &[P_H(\pi/4,0)]_{\text{D}}=\frac{1}{4}(I_{HH}+I_{VH}+1)+\frac{1}{2}\sqrt{I_{VH}}J_{VH}^{VV}\sin(\phi-(\phi_p)_{\text{D}}+\phi^{VV}_{VH})\label{p-h-4-phase-2},\\
    &[P_V(0,\pi/4)]_{\text{D}}=\frac{1}{4}(I_{HV}+I_{VV}+1)+\frac{1}{2}\sqrt{I_{HV}}J_{HV}^{VV}\sin(\phi-(\phi_p)_{\text{D}}+\phi_{HV}^{VV})\label{p-v-4-phase-1},\\
    &[P_V(\pi/4,\pi/4)]_{\text{D}}=\frac{1}{4}(I_{HV}+I_{VV}+1)+\frac{1}{2}\sqrt{I_{VV}}\sin(\phi-(\phi_p)_{\text{D}})\label{p-v-4-phase-2}.
	\end{align}
\end{subequations}
\end{widetext}
\par
Let us now define
\begin{align}\label{Ppm}
    [P_\mu^{(-)}(\theta,\delta)]_j=\left\{[P_\mu(\theta,\delta)]_j\right\}_{\text{max}}-\left\{[P_\mu(\theta,\delta)]_j\right\}_{\text{min}},
\end{align}
where $\mu=H, V$ and the maximum and minimum values of $[P_\mu(\theta,\delta)]_j$ can be obtained by varying the phase $\phi$. Using Eqs.~(\ref{q-state}), (\ref{p-h-1-phase-1})-(\ref{p-v-4-phase-2}) and (\ref{Ppm}), we obtain all diagonal elements and moduli of all off-diagonal elements of the unknown density matrix ($\dm$) as follows:
\begin{widetext}
    \begin{subequations}
    \begin{align}
        &\bra{H_u H_d} \,\dm \,\ket{H_u H_d}=I_{HH}=\left\{[P_H^{(-)}(0,0)]_A\right\}^2\label{I-11},\\
        &\bra{H_u V_d} \,\dm \,\ket{H_u V_d}=I_{HV}=\left\{[P_V^{(-)}(0,\pi/4)]_B\right\}^2\label{I-22},\\
        &\bra{V_u H_d} \,\dm \,\ket{V_u H_d}=I_{VH}=\left\{[P_H^{(-)}(\pi/4,0)]_C\right\}^2\label{I-33},\\
        &\bra{V_u V_d} \,\dm \,\ket{V_u V_d}=I_{VV}=\left\{[P_V^{(-)}(\pi/4,\pi/4)]_D\right\}^2\label{I-44},\\
        &|\bra{H_u H_d} \,\dm \,\ket{H_u V_d}|=\sqrt{I_{HH}I_{HV}}J_{HV}^{HH}=[P_H^{(-)}(0,0)]_A [P_V^{(-)}(0,\pi/4)]_A=[P_V^{(-)}(0,\pi/4)]_B[P_H^{(-)}(0,0)]_B\label{d-12},\\
        &|\bra{H_u H_d} \,\dm \,\ket{V_u H_d}|=\sqrt{I_{HH}I_{VH}}J_{VH}^{HH}=[P_H^{(-)}(0,0)]_A [P_H^{(-)}(\pi/4,0)]_A=[P_H^{(-)}(\pi/4,0)]_C[P_H^{(-)}(0,0)]_C\label{d-13},\\
        &|\bra{H_u H_d} \,\dm \,\ket{V_u V_d}|=\sqrt{I_{HH}I_{VV}}J_{VV}^{HH}=[P_H^{(-)}(0,0)]_A [P_V^{(-)}(\pi/4,\pi/4)]_A=[P_V^{(-)}(\pi/4,\pi/4)]_D[P_H^{(-)}(0,0)]_D\label{d-14},\\
        &|\bra{H_u V_d} \,\dm \,\ket{V_u H_d}|=\sqrt{I_{HV}I_{VH}}J_{VH}^{HV}=[P_V^{(-)}(0,\pi/4)]_B[P_H^{(-)}(\pi/4,0)]_B=[P_H^{(-)}(\pi/4,0)]_C[P_V^{(-)}(0,\pi/4)]_C\label{d-23},\\
        &|\bra{H_u V_d} \,\dm \,\ket{V_u V_d}|=\sqrt{I_{HV}I_{VV}}J_{VV}^{HV}=[P_V^{(-)}(0,\pi/4)]_B[P_V^{(-)}(\pi/4,\pi/4)]_B=[P_V^{(-)}(\pi/4,\pi/4)]_D[P_V^{(-)}(0,\pi/4)]_D\label{d-24},\\
        &|\bra{V_u H_d} \,\dm \,\ket{V_u V_d}|=\sqrt{I_{VH}I_{VV}}J_{VV}^{VH}=[P_H^{(-)}(\pi/4,0)]_C[P_V^{(-)}(\pi/4,\pi/4)]_C=[P_V^{(-)}(\pi/4,\pi/4)]_D[P_H^{(-)}(\pi/4,0)]_D\label{d-34}.
    \end{align}
\end{subequations}
\end{widetext}
 We note that the modulus of each off-diagonal element can be determined equivalently using two configurations. For example, Eq.~(\ref{d-14}) shows that the modulus of $|\bra{H_u H_d} \,\dm \,\ket{V_u V_d}|$ can be determined using both configurations A and D. 
\par
We now show how to determine the arguments (phases) of the off-diagonal elements. These arguments are given by phase parameters since $\text{arg}\{\bra{\mu_u \nu_d} \,\dm \,\ket{\mu'_u \nu'_d}\}=\phi_{\mu\nu}^{\mu'\nu'}$. Let us first observe from Eqs.~(\ref{p-h-1-phase-1}) and (\ref{p-v-1-phase-2}) that the phase parameter $\phi_{VV}^{HH}=\text{arg}\{\bra{V_u V_d} \,\dm \,\ket{H_u H_d}\}$ is equal to the phase difference between the interference patterns $[P_H(0,0)]_{\text{A}}$ and $[P_V(\pi/4,\pi/4)]_{\text{A}}$. Therefore, $\phi_{VV}^{HH}$ can be determined by comparing these two interference patterns obtained in configuration A. We also note from Eqs.~(\ref{p-h-4-phase-1}) and (\ref{p-v-4-phase-2}) that $\phi_{VV}^{HH}$ can alternatively be determined using the interference patterns $[P_V(\pi/4,\pi/4)]_{\text{D}}$ and $[P_H(0,0)]_{\text{D}}$ obtained in configuration D. In Fig.~\ref{Bell-D}, we illustrate how $\phi_{VV}^{HH}$ can be retrieved by considering a two-qubit generalized Bell state. Arguments of the remaining off-diagonal elements of $\dm$ can be determined following the same procedure. Each of them can be determined using two alternative configurations, identical to those used to obtain the corresponding modulus, as evident from Eqs.~(\ref{p-h-1-phase-1})-(\ref{p-v-4-phase-2}) and Eqs.~(\ref{d-12})-(\ref{d-34}). In Table~\ref{phase-tab-res}, we list the interference patterns and corresponding configurations which can be used to determine arguments of all off-diagonal elements. 
\begin{table*}[t]		
	\setlength{\tabcolsep}{5 pt} 
	\renewcommand{\arraystretch}{1} 
	\begin{tabular}{ccc}
		\hline\hline 
		Arguments of density& \multirow{2}*{Configurations} & \multirow{2}*{Interference patterns}\\  matrix elements &\\[2 pt] 
		\hline 
		$\text{arg}\{\bra{H_u V_d} \,\dm \,\ket{H_u H_d}\}$ & A or B & \{$[P_H(0,0)]_{\text{A}}$,$[P_V(0,\pi/4)]_{\text{A}}$\} or \{$[P_V(0,\pi/4)]_{\text{B}}$,$[P_H(0,0)]_{\text{B}}$\}\\[1ex] 
		$\text{arg}\{\bra{V_u H_d} \,\dm \,\ket{H_u H_d}\}$ & A or C & $\{[P_H(0,0)]_{\text{A}}$,$[P_H(\pi/4,0)]_{\text{A}}\}$ or $\{[P_H(\pi/4,0)]_{\text{C}}$,$[P_H(0,0)]_{\text{C}}\}$ \\[1ex] 
		$\text{arg}\{\bra{V_u V_d} \,\dm \,\ket{H_u H_d}\}$ & A or D & $\{[P_H(0,0)]_{\text{A}}$, $[P_V(\pi/4,\pi/4)]_{\text{A}}\}$ or $\{[P_V(\pi/4,\pi/4)]_{\text{D}}$, $[P_H(0,0)]_{\text{D}}\}$  \\[1ex] 
        $\text{arg}\{\bra{V_u H_d} \,\dm \,\ket{H_u V_d}\}$ & B or C &  $\{[P_V(0,\pi/4)]_{\text{B}}$,$[P_H(\pi/4,0)]_{\text{B}}\}$ or $\{[P_H(\pi/4,0)]_{\text{C}}$,$[P_V(0,\pi/4)]_{\text{C}}\}$ \\[1ex] 
        $\text{arg}\{\bra{V_u V_d} \,\dm \,\ket{H_u V_d}\}$ & B or D & $\{[P_V(0,\pi/4)]_{\text{B}}$, $[P_V(\pi/4,\pi/4)]_{\text{B}}\}$ or $\{[P_V(\pi/4,\pi/4)]_{\text{D}}$, $[P_V(0,\pi/4)]_{\text{D}}\}$\\[1ex] 
        $\text{arg}\{\bra{V_u H_d} \,\dm \,\ket{V_u V_d}\}$& C or D & $\{[P_H(\pi/4,0)]_{\text{C}}$, $[P_V(\pi/4,\pi/4)]_{\text{C}}\}$ or $\{[P_V(\pi/4,\pi/4)]_{\text{D}}$, $[P_H(\pi/4,0)]_{\text{D}}\}$  \\[1ex] 
		\hline\hline
	\end{tabular}
	\caption{Interference patterns that can be used for determining arguments of off-diagonal density matrix elements.}\label{phase-tab-res}
\end{table*}
\par
\begin{figure}[htbp]
\centering
\includegraphics[width=\linewidth]{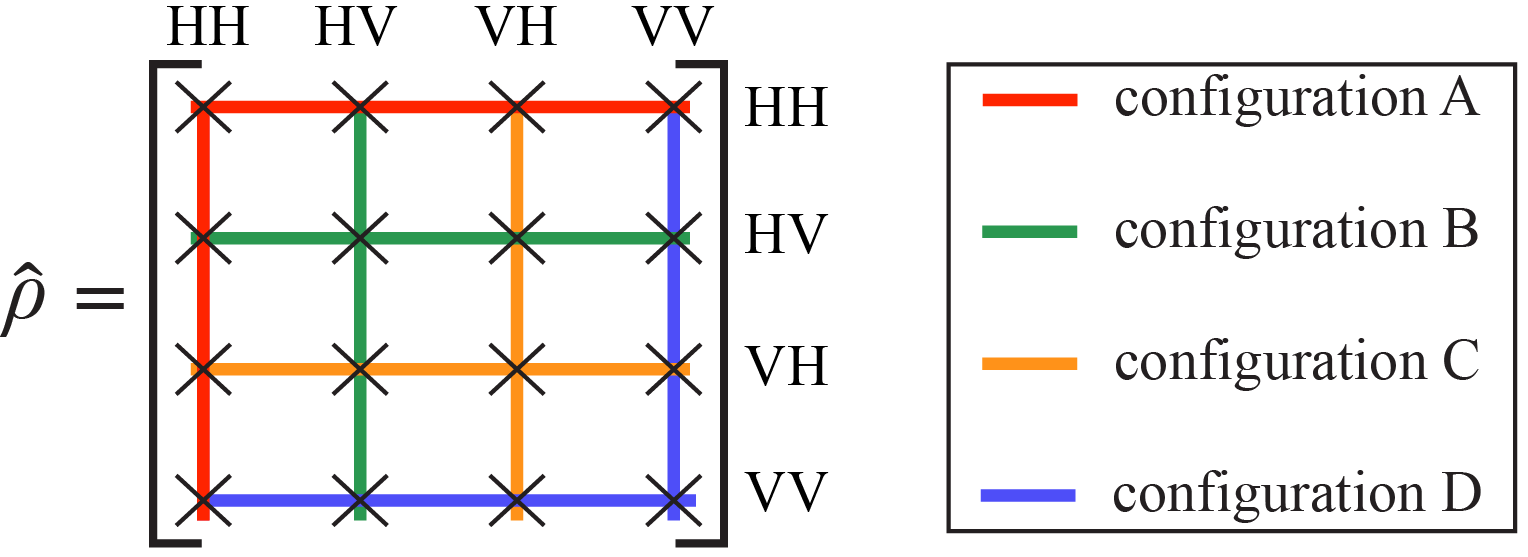}
\caption{Configurations (solid lines) corresponding to each density matrix element (cross marks).}
\label{fig:state-determination}
\end{figure}

It can be noted that each diagonal element of $\dm$ is obtained using a distinct configuration, whereas each off-diagonal element can be obtained using two alternative configurations. This is fully consistent with the fact that $\dm$ is a Hermitian matrix, i.e., the diagonal elements are real and the off-diagonal elements are related by $\bra{\mu_u \nu_d} \,\dm \,\ket{\mu'_u \nu'_d}=\bra{\mu'_u \nu'_d} \,\dm \,\ket{\mu_u \nu_d}^{\ast}$. Figure \ref{fig:state-determination} depicts the configurations that can be used to determine each element of the density matrix. In this figure, each configuration is represented by a line of distinct color, and each matrix element is represented by a cross mark. This figure also identifies the matrix elements that can be found in each configuration. For example, the red line along the first row and the first column of the matrix implies that the following matrix elements are determinable in configuration A: $\bra{H_u H_d} \,\dm \,\ket{H_u H_d}$, $\bra{H_u H_d} \,\dm \,\ket{H_u V_d}$, $\bra{H_u H_d} \,\dm \,\ket{V_u H_d}$, $\bra{H_u H_d} \,\dm \,\ket{V_u V_d}$, $\bra{H_u V_d} \,\dm \,\ket{H_u H_d}$, $\bra{V_u H_d} \,\dm \,\ket{H_u H_d}$, and $\bra{V_u V_d} \,\dm \,\ket{H_u H_d}$. 
\par
Equations~(\ref{I-11})-(\ref{d-34}) and Table~\ref{phase-tab-res} contain our main results: the diagonal elements of the density matrix ($\dm$) are given by Eqs.~(\ref{I-11})-(\ref{I-44}). The off-diagonal elements are given by their magnitudes from Eqs.~(\ref{d-12})-(\ref{d-34}) and their arguments (phases) from Table~\ref{phase-tab-res}. 

\section{Illustration}\label{sec:illustration}
We now illustrate our results by 
considering a mixed state of the form (see Fig.~\ref{notation-1}b for the matrix form)
\begin{align}\label{gen-bell-state}  \dm&=I_{HH}\ket{H_uH_d}\bra{H_uH_d}+I_{VV}\ket{V_uV_d}\bra{V_uV_d} \nonumber \\ &+\sqrt{I_{HH}I_{VV}}J_{HH}^{VV} e^{-i\phi_{VV}^{HH}} \ket{H_uH_d}\bra{V_uV_d} \nonumber\\ &+ \sqrt{I_{HH}I_{VV}} J_{HH}^{VV}e^{i\phi_{VV}^{HH}} \ket{V_uV_d}\bra{H_uH_d}.
\end{align}
This density operator contains three free parameters: $I_{HH}, J_{HH}^{VV}$ and $\phi_{VV}^{HH}$, with $I_{VV}=1-I_{HH}$. The state given by Eq.~(\ref{gen-bell-state}) is obtained by generalizing the Bell state $(\ket{H_uH_d}+\ket{V_uV_d})/\sqrt{2}$, and therefore, we call it a two-qubit generalized Bell state \cite{gen-Bell-note}.
\par
It follows from Eqs.~(\ref{p-h-1-phase-1})-(\ref{p-v-4-phase-2}) that in this case, single-photon interference patterns with non-zero visibility can be obtained in configurations A and D only. That is, in an experiment one would find only these two configurations (see also Appendix \ref{app:finding-configs}). Therefore, one can readily conclude that all density matrix elements covered by configurations B and C (Fig.~\ref{fig:state-determination}) must be zero \cite{nonzero-vis}. 
\par
To illustrate retrieval of non-zero matrix elements, we consider the following density matrix
\begin{align}\label{Bell-state-matrix}
	\setlength{\arraycolsep}{0.5cm}
	\begin{pmatrix}
	0.4 & 0 & 0 &-0.3-i0.3 \\
	0 & 0 & 0 & 0 \\
	0 & 0 & 0 & 0 \\
	-0.3+i0.3 & 0 & 0 & 0.6
	\end{pmatrix}.
	\end{align} 
In this case, the four relevant interference patterns are given by Eqs.~(\ref{p-h-1-phase-1}), (\ref{p-v-1-phase-2}), (\ref{p-h-4-phase-1}), and (\ref{p-v-4-phase-2}), which take the form
\begin{subequations}\label{interf-example}
\begin{align}
&[P_H(0,0)]_{\text{A}}\approx 0.35+0.32\sin(\phi-(\phi_p)_{\text{A}}) , \label{interf-example:a} \\
&\left[P_V\left(\pi/4,\pi/4\right)\right]_{\text{A}}\approx 0.4+0.34\sin(\phi-(\phi_p)_{\text{A}}+2.36) , \label{interf-example:b} \\ 
&[P_H(0,0)]_{\text{D}}\approx 0.35+0.28 \sin(\phi-(\phi_p)_{\text{D}}-2.36), \label{interf-example:} \\
&\left[P_V\left(\pi/4,\pi/4\right)\right]_{\text{D}}\approx 0.4+0.39\sin(\phi-(\phi_p)_{\text{D}}). \label{interf-example:d}
\end{align}
\end{subequations}
\begin{figure}[htbp]
	\centering
	\includegraphics[width=\linewidth]{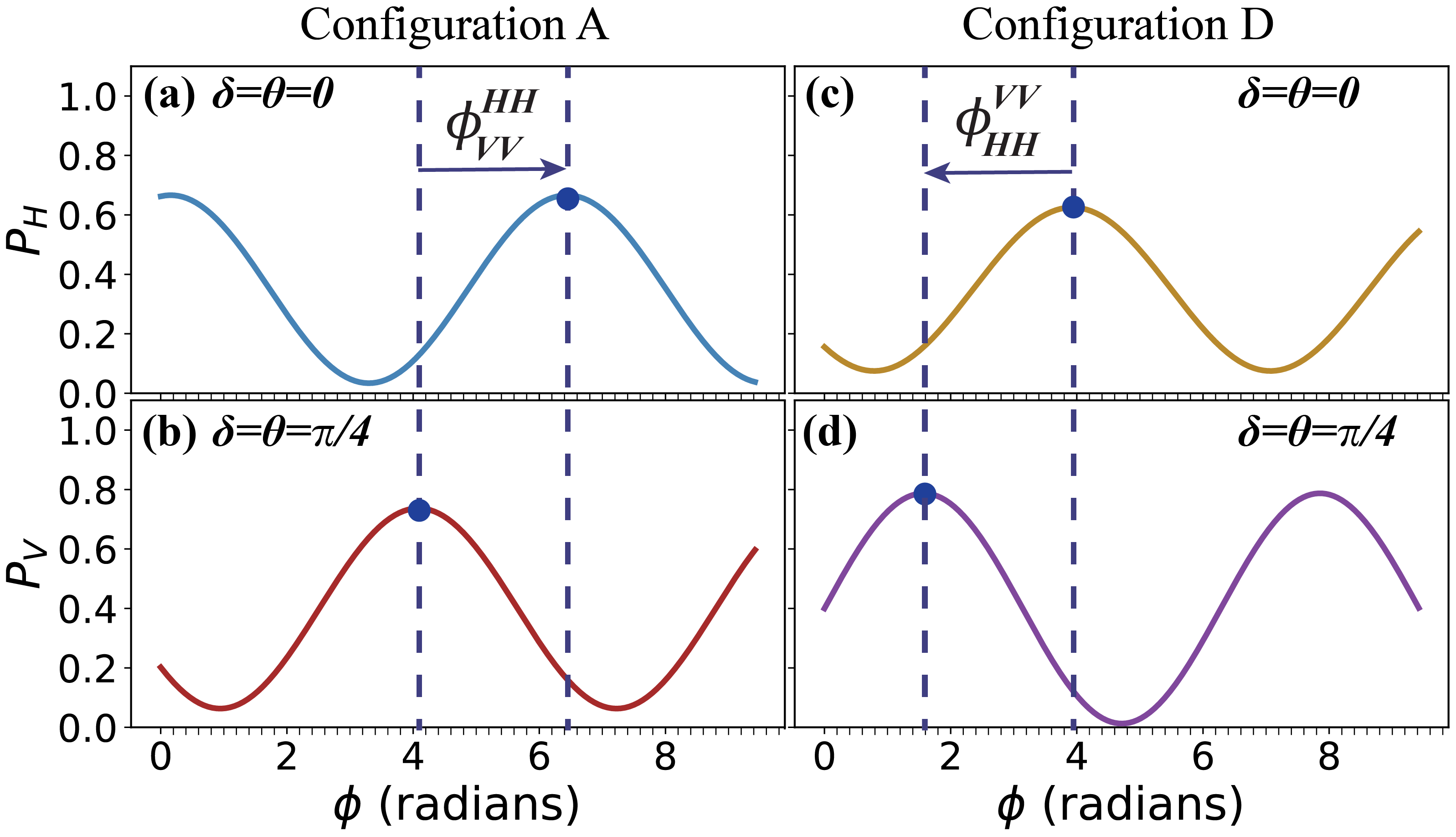}
	\qquad
	\caption{Determining arguments of the off-diagonal density matrix elements for the two-qubit  generalized Bell state given by Eq.~(\ref{Bell-state-matrix}). We have chosen $(\phi_p)_{\text{A}}=0.45\pi$ and $(\phi_p)_{\text{D}}=0$. Retrieved arguments are independent of the choices of $(\phi_p)_{\text{A}}$ and $(\phi_p)_{\text{D}}$. (a) The phase difference between interference patterns $[P_H(0,0)]_{\text{A}}$ and $[P_V(\pi/4,\pi/4)]_{\text{A}}$ gives $\phi_{VV}^{HH}=\text{arg}\{\bra{V_u V_d} \,\dm \,\ket{H_u H_d}\}=2.36$. (b) The phase difference between interference patterns $[P_V(\pi/4,\pi/4)]_{\text{D}}$ and $[P_H(0,0)]_{\text{D}}$ gives $\phi_{HH}^{VV}=\text{arg}\{\bra{H_u H_d} \,\dm \,\ket{V_u V_d}\}=-2.36$.}\label{Bell-D}
\end{figure}
The interference patterns given by Eqs.~(\ref{interf-example:a})--(\ref{interf-example:d}) are illustrated by Figs.~\ref{Bell-D}a--\ref{Bell-D}d, respectively, where the probability of detecting a $d$-photon is plotted against the experimentally tunable phase $\phi$.  
\par
\begin{figure}[htbp]
	\centering
	\includegraphics[width=\linewidth]{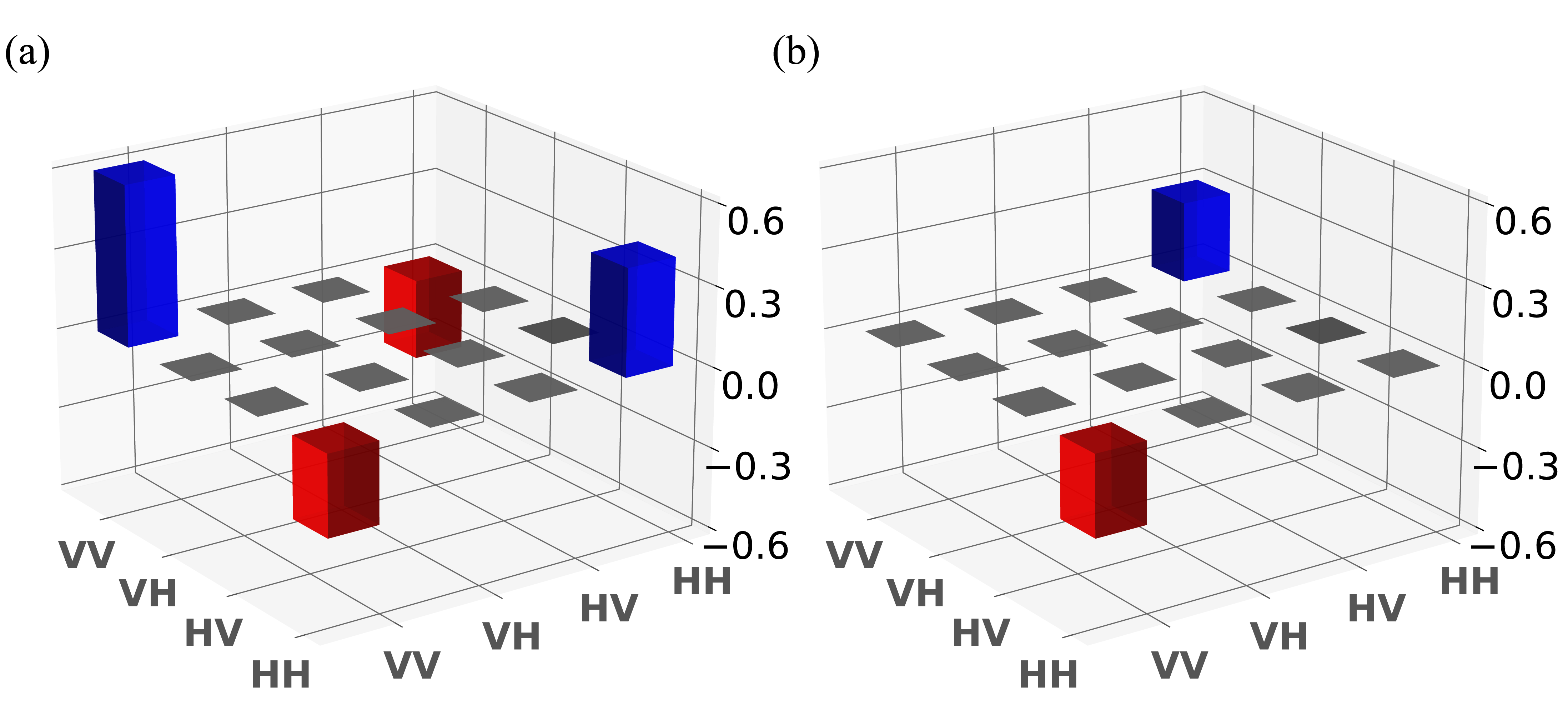}
	\qquad
	\caption{Real parts (a) and imaginary parts (b) of retrieved density matrix elements of the two-qubit  generalized Bell state given by Eq.~(\ref{Bell-state-matrix}).}\label{fig:illustration-figure}
\end{figure}

The non-zero diagonal elements are obtained using Eqs.~(\ref{I-11}) and (\ref{I-44}). The magnitudes of the two non-zero off-diagonal elements are obtained using Eq.~(\ref{d-14}). Arguments (phases) of the non-zero off-diagonal elements are obtained by comparing interference patterns $[P_H(0,0)]_{\text{A}}$ and $[P_V(\pi/4,\pi/4)]_{\text{A}}$, or equivalently, by comparing $[P_H(0,0)]_{\text{D}}$ and $[P_V(\pi/4,\pi/4)]_{\text{D}}$, as illustrated by Fig.~\ref{Bell-D}. The retrieved density matrix elements are in excellent agreement with the values given in Eq.~(\ref{Bell-state-matrix}). The results are illustrated by Fig.~\ref{fig:illustration-figure}.

\section{Treating key Experimental Imperfections}\label{sec:loss-main}
Although the objective of a theoretical paper is not to address possible experimental imperfections, one anticipated imperfection in the proposed experiment deserves theoretical attention. In two related experiments \textemdash ~entanglement measurement in a two-qubit system without detecting one qubit \cite{lemos2023one} and single-qubit state tomography without detecting the qubit \cite{fuenzalida2024quantum} \textemdash ~misalignment of undetected photon beams and loss of undetected photons between the sources emerged as the dominant imperfections requiring careful consideration. This is because these imperfections reduce interference visibility that is also dependent on certain state parameters. This loss of visibility cannot be compensated for and consequently, a quantitative analysis of their impact on the experimental data is essential.
\par
Theoretical analysis \cite{lahiri2021characterizing} supported by experimental observation \cite{lemos2023one} have demonstrated that these imperfections can be theoretically modeled by introducing an attenuator with polarization-dependent transmittance. In our case (Fig.~\ref{fig:setup}), since source $Q_2$ does not generate any $V$-polarized photons, the loss of undetected $V$-polarized photons in the beam $u_1$ is irrelevant. Therefore, an attenuator that only attenuates $H$-polarized photons is sufficient for our purpose. In Appendix~\ref{app:gen-treatment-loss}, we provide a detailed analysis showing how to quantitatively treat these imperfections and reconstruct the density matrix elements in their presence. 

\section{Discussion and Outlook}\label{sec:discuss}
We have presented a method to fully reconstruct the density matrix of a two-qubit system without detecting one of the qubits. Our method is applicable to both pure and mixed qubit states, including the most general one that is characterized by $15$ free parameters. Since our analysis is based on quantum field theory, the method is applicable to any non-photonic qubit for which similar interferometers can be constructed. 
\par
A complete characterization of the density matrix allows for the determination of any entanglement measure. Therefore, our work can be viewed as a generalization of recent approaches to measure entanglement in bipartite qubit systems without directly measuring one of the subsystems. Prior work on this topic has addressed entanglement measurement for two-qubit generalized Bell \cite{lahiri2021characterizing,lemos2023one} and Werner states \cite{rajeev2023single}. 
In this context, the experiment presented in Ref.~\cite{lemos2023one} has demonstrated the feasibility of using path-identity interferometers for quantum state measurement. In Ref.~\cite{lemos2023one}, two identical sources were used each generating the same quantum state, whereas the present method employs two distinct sources: one to generate the unknown state and the other to generate a known reference state. The use of distinct sources in a path-identity interferometer has also been experimentally implemented while performing single-photon state tomography without detecting the photon \cite{fuenzalida2024quantum}. These two recent experiments \cite{lemos2023one,fuenzalida2024quantum}, therefore, strongly suggest that our method is experimentally implementable.
\par
While our results show that all information contained in an arbitrary two-qubit photonic state can be retrieved without detecting one of the qubits, it remains to be answered whether this holds for any high-dimensional two-photon state. A proposal for characterizing entanglement in mixed qudit Bell states using path-identity interferometers has been presented in Ref.~\cite{zhan2021determining}. There have also been experimental implementations of path-identity interferometers with photonic two-qudit states \cite{kysela2020path}. We, therefore, hope that our work will inspire further exploration of high-dimensional photonic systems through the lens of path identity. 
\par
Finally, beyond answering the fundamental question of whether complete state tomography of an arbitrary two-qubit state is possible by detecting one qubit only, our findings offer practical advantages in certain scenarios. Detection of low-intensity light is challenging for a wide spectral range. For example, while good single-photon detectors are readily available for visible and near-infrared light, effective single-photon detectors for mid- and far-infrared regions are not easily accessible \cite{lei2015progress,verma2021single}. Our method circumvents the need to detect both photons, offering a unique advantage over existing techniques when one of the two photons is challenging or impossible to detect. Our approach thus enables the measurement of two-photon states that are currently inaccessible due to lack of effective detectors.

\section*{Acknowledgments}
The research was supported by the US Air Force Office of Scientific Research under grant FA9550-23-1-0216.

\appendix

\section{Matrix form of $\dm$}\label{app:matrix-dm}
An arbitrary two-qubit mixed state, given by Eq.~(\ref{q-state}) in the main text, takes the following matrix form in the computational basis $\{\ket{H_uH_d}, \ket{H_uV_d}, \ket{V_uH_d}, \ket{V_uV_d}\}$:
\begin{widetext}
	\begin{align}\label{q-state-matrix}
	\setlength{\arraycolsep}{0.5cm}
	\dm=\begin{pmatrix}
	I_{HH} & \sqrt{I_{HH}I_{HV}}J_{HH}^{HV}e^{i\phi_{HH}^{HV}} & \sqrt{I_{HH}I_{VH}}J_{HH}^{VH}e^{i\phi_{HH}^{VH}} & \sqrt{I_{HH}I_{VV}}J_{HH}^{VV}e^{i\phi_{HH}^{VV}} \\[7pt]
	\sqrt{I_{HH}I_{HV}}J_{HV}^{HH}e^{-i\phi_{HH}^{HV}} & I_{HV} & \sqrt{I_{VH}I_{HV}}J_{HV}^{VH}e^{i\phi_{HV}^{VH}} & \sqrt{I_{VV}I_{HV}}J_{HV}^{VV}e^{i\phi_{HV}^{VV}}\\[7pt]
	\sqrt{I_{HH}I_{VH}}J_{HH}^{VH}e^{-i\phi_{HH}^{VH}} & \sqrt{I_{VH}I_{HV}}J_{HV}^{VH}e^{-i\phi_{HV}^{VH}} & I_{VH} & \sqrt{I_{VH}I_{VV}}J_{VH}^{VV}e^{i\phi_{VH}^{VV}}\\[7pt]
	\sqrt{I_{HH}I_{VV}}J_{HH}^{VV}e^{-i\phi_{HH}^{VV}} & \sqrt{I_{VV}I_{HV}}J_{HV}^{VV}e^{-i\phi_{HV}^{VV}} & \sqrt{I_{VH}I_{VV}}J_{VH}^{VV}e^{-i\phi_{VH}^{VV}} & I_{VV}
	\end{pmatrix}.
	\end{align} 
\end{widetext}

\section{General treatment considering unequal emission probabilities from sources and imperfect alignment}\label{app:gen-treatment-loss}

\subsection{Unequal emission probabilities}\label{app:total-density-matrix}

Here, we consider a scenario in which sources $Q_1$ and $Q_2$ may not emit with equal probabilities. Following the same approach to obtain Eq.~(\ref{q-state-total}) in the main text, we find that the state generated by coherent emissions of $Q_1$ and $Q_2$ is given by
\begin{align}\label{q-state-total-b1b2}
&(\dmt_{du})_j=|b_1|^2\sum_{\mu,\nu}^{H,V} \sum_{\mu',\nu'}^{H,V} \sqrt{I_{\mu\nu}I_{\mu'\nu'}}J_{\mu\nu}^{\mu'\nu'}\exp\{i\phi_{\mu\nu}^{\mu'\nu'}\} \nonumber\\&\times\ket{\mu_{u_1}\nu_{d_1}} \bra {\mu'_{u_1}\nu'_{d_1}}+|b_2|^2\ket{H_{u_2}H_{d_2}}\bra{H_{u_2}H_{d_2}}\nonumber\\&+\big\{b_1b_2^*\sum_{\mu,\nu}^{H,V} \sqrt{I_{\mu\nu}}(J_{\mu_1\nu_1}^{H_2H_2})_j\exp\{{i(\phi_{\mu_1\nu_1}^{H_2H_2}})_j\} \nonumber\\&\times\ket{\mu_{u_1}\nu_{d_1}} \bra{H_{u_2}H_{d_2}}+H.c.\big\},
\end{align}
where $j=\text{A},\text{B},\text{C},\text{D}$ denotes a configuration, $H.c.$ denotes Hermitian conjugation, $|b_k|^2 (k=1,2)$ is proportional to the probability of emission from source $Q_k$, such that $|b_1|^2+|b_2|^2=1$; the parameters $I_{\mu\nu}$, $J_{\mu\nu}^{\mu'\nu'}$, $J_{\mu_1\nu_1}^{H_2H_2}$, $\phi_{\mu\nu}^{\mu'\nu'}$ and $\phi_{\mu_1\nu_1}^{H_2H_2}$ are defined in the main text. We observe that when $|b_1|=|b_2|=1/\sqrt{2}$ and $\text{arg}\{b_1\}-\text{arg}\{b_2\}=\phi_0$, Eq. (\ref{q-state-total-b1b2}) reduces to Eq.~(\ref{q-state-total}) of the main text.

\subsection{Path identity under experimental imperfections}\label{app:alignment-conditions-loss}
\begin{figure}[htbp]  \centering
\includegraphics[width=\linewidth]{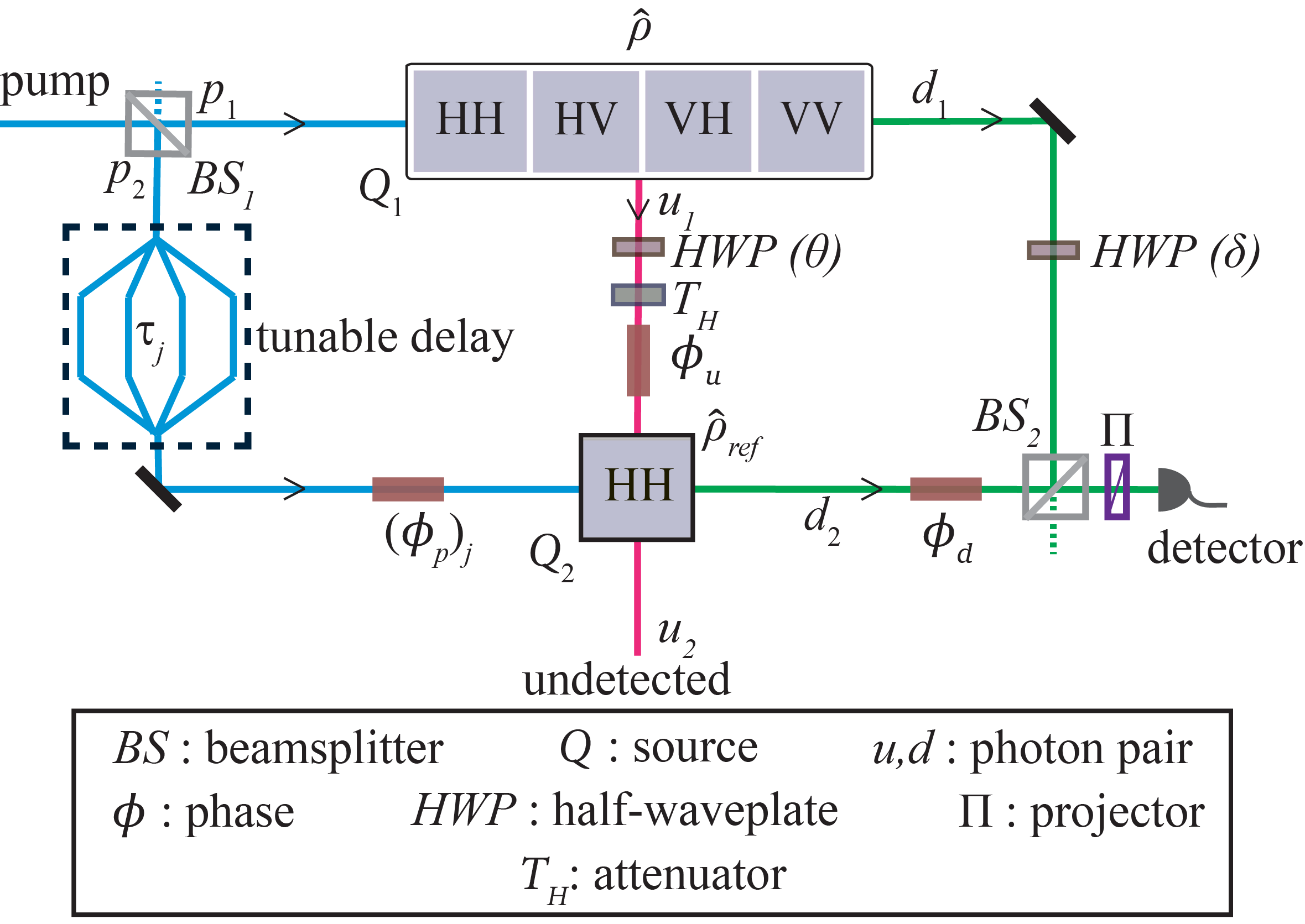}
\qquad
\caption{The dominant experimental imperfections are modeled by introducing a filter ($T_H$) that attenuates $H$-polarized $u$-photons in beam $u_1$. The rest of the setup is the same as that in Fig.~\ref{fig:setup}.} \label{fig:scheme-supp}
\end{figure}

As discussed in Sec.~\ref{sec:loss-main} of the main text, the key experimental imperfections which require careful consideration are the misalignment of undetected photon beams and loss of undetected photons between the sources. These imperfections can be effectively modeled by introducing an attenuator with polarization dependent transmittance in beam $u_1$ between the half-wave plate and $Q_2$ (see Fig.~\ref{fig:scheme-supp}). In our case, since source $Q_2$ does not generate any $V$-polarized photons, the loss of undetected $V$-polarized photons in the beam $u_1$ is irrelevant. Therefore, an attenuator that only attenuates $H$-polarized photons is sufficient for our purpose.
\par
The action of the attenuator is mathematically similar to that of beamsplitter ($BS$) with single input \cite{zou1991induced}. We denote the amplitude transmission coefficient for $H$-polarized photon as $T_H$. We can assume, without loss of generality, that $T_H$ is real and $0\leq T_H\leq1$. 
\par
Let us now observe that a $u$-photon in beam $u_1$ first undergoes the transformation $HWP(\theta)$, then passes through the attenuator, and finally enters beam $u_2$ through source $Q_2$ (Fig.~\ref{fig:scheme-supp}). Consequently, the quantum fields corresponding to the $u$-photon at $Q_1$ and $Q_2$ are related by
\begin{align} \label{align-cond-field-supp}
&\hat{a}_{u_2}(H)=e^{i\phi_u}\Big[T_H\big\{\hat{a}_{u_1}(H) \cos2\theta+\hat{a}_{u_1}(V) \sin2\theta\big\} \nonumber \\ & \qquad \quad+R_H~\hat{a}_{0}(H)\Big],
\end{align} 
where $R_H=\sqrt{1-(T_H)^2}$, $\hat{a}_{u_j}(\mu)$ represents the annihilation operator corresponding to the undetected photon in state $\ket{\mu_{u_j}}$, and $\phi_u$ represents the phase change due to propagation from $Q_1$ to $Q_2$. Using $\hat{a}^{\dag}_{u_j}(\mu)\ket{\text{vac}}=\ket{\mu_{u_j}}$, Eq.~(\ref{align-cond-field-supp}) can also be expressed as
		\begin{align} \label{align-cond-kets-supp}
		\ket{H_{u_2}}&=e^{-i\phi_u} \big[T_H \big(\ket{H_{u_1}}\cos 2\theta + \ket{V_{u_1}} \sin 2\theta \big)  \nonumber \\ & \qquad \quad+ R_H\ket{H_0}\big], 
		\end{align} 
where $\ket{H_0}=\hat{a}^{\dag}_{0}(H)\ket{\text{vac}}$ can be interpreted as the state of a lost $u$-photon with polarization $H$. One can check that in the ideal scenario, where $T_H=1$ (i.e., $R_H=0$), Eqs.~(\ref{align-cond-field-supp}) and (\ref{align-cond-kets-supp})
reduce to Eqs.~(\ref{algn-operats}) and (\ref{algn}) in the main text, respectively.

\subsection{Determining the density operator considering unequal emission probabilities and imperfect alignment}\label{app:reduced-density-operator}
Substituting for $\ket{H_{u_2}}$ from Eq.~(\ref{align-cond-kets-supp}) into Eq. (\ref{q-state-total-b1b2}), we find that the two-photon density operator in configuration $j=\text{A},\text{B},\text{C},\text{D}$ is given by
\begin{widetext}
    \begin{align}\label{q-state-total-algn-supp}
	&(\dm_{du})_j=|b_1|^2\sum_{\mu,\nu}^{H,V} \sum_{\mu',\nu'}^{H,V} \sqrt{I_{\mu\nu}I_{\mu'\nu'}}\exp\big(i\phi_{\mu\nu}^{\mu'\nu'}\big) \ket{\mu_{u_1}\nu_{d_1}}\bra {\mu'_{u_1}\nu'_{d_1}}+|b_2|^2\Big(\sum_{\lambda,\lambda'}^{H,V}T_H^2[\hat{O}_1]_{H\lambda}[\hat{O}_1]_{H\lambda'}\ket{\lambda_{u_1}H_{d_2}}\bra{\lambda'_{u_1}H_{d_2}}\nonumber\\&+T_H R_H\Big[\sum_{\lambda}^{H,V}[\hat{O}_1]_{H\lambda}\ket{\lambda_{u_1}H_{d_2}}\bra{H_0H_{d_2}}+\sum_{\lambda}^{H,V}[\hat{O}_1]_{H\lambda'}\ket{H_0H_{d_2}}\bra{\lambda'_{u_1}H_{d_2}}\Big]+R_H^2\ket{H_0H_{d_2}}\bra{H_0H_{d_2}}\Big) \nonumber \\ &+\Big\{b_1b_2^*e^{i\phi_u} \Big[ \sum_{\mu,\nu}^{H,V} \sqrt{I_{\mu\nu}}(J_{\mu_1\nu_1}^{H_2H_2})_j\exp\left[i(\phi_{\mu_1\nu_1}^{H_2H_2})_j \right]\Big( T_H\sum_{\lambda}[\hat{O}_1]_{H\lambda}\ket{\mu_{u_1}\nu_{d_1}} \bra{\lambda_{u_1}H_{d_2}}+R_H\ket{\mu_{u_1}\nu_{d_1}} \bra{H_0H_{d_2}} \Big)\Big]+H.c.\Big\},
	\end{align}
    \end{widetext}
where $[\hat{O}_1]_{HH}=\cos2\theta$ and $[\hat{O}_1]_{HV}=\sin2\theta$.
\par
The reduced density matrix, $(\dm_d)_j$, representing a $d$-photon is obtained by taking the partial trace of $(\dm_{du})_j$ over the subspace complementary to the $d$-photon subspace. That is,
\begin{align}\label{reduced-density-def}
    (\dm_d)_{j}=\sum_{\lambda_{u_1}}^{H,V}\bra{\lambda_{u_1}}(\dm_{du})_j\ket{\lambda_{u_1}}+\bra{H_0}(\dm_{du})_j\ket{H_0}.
\end{align}
Using Eqs.~(\ref{q-state-total-algn-supp}) and (\ref{reduced-density-def}), we obtain
\begin{widetext}
\begin{align}\label{signal-dm-supp} 
&(\dm_d)_j=|b_1|^2\Big(\sum_{\mu,\nu}^{H,V}I_{\mu\nu}\ket{\nu_{d_1}}\bra{\nu_{d_1}}+\Big\{\big(\sqrt{I_{HH}I_{HV}}\I_{HH}^{HV}e^{i\phi_{HH}^{HV}}+\sqrt{I_{VH}I_{VV}}\I_{VH}^{VV}e^{i\phi_{VH}^{VV}}\big)\ket{H_{d_1}}\bra{V_{d_1}}+H.c.\Big\}\Big)\nonumber\\&+|b_2|^2\ket{H_{d_2}}\bra{H_{d_2}}+\Big[b_1b_2^*e^{i\phi_u}T_H\big(\cos 2\theta\big\{\sqrt{I_{HH}}(J_{H_1H_1}^{H_2H_2} )_j e^{i(\phi_{H_1H_1}^{H_2H_2})_j}\ket{H_{d_1}}\bra{H_{d_2}}+\sqrt{I_{HV}}(J_{H_1V_1}^{H_2H_2})_j e^{i(\phi_{H_1V_1}^{H_2H_2})_j}\nonumber\\&\times \ket{V_{d_1}}\bra{H_{d_2}}\big\}+\sin 2\theta\big\{\sqrt{I_{VH}}(J_{V_1H_1}^{H_2H_2})_j e^{i(\phi_{V_1H_1}^{H_2H_2})_j}\ket{H_{d_1}}\bra{H_{d_2}}+\sqrt{I_{VV}}(J_{V_1V_1}^{H_2H_2})_j e^{i(\phi_{V_1V_1}^{H_2H_2})_j}\ket{V_{d_1}}\bra{H_{d_2}}\big\}\big)+H.c.\Big].
\end{align}
\end{widetext}
Equation~(\ref{signal-dm-supp}) reduces to Eq.~(\ref{detected-dm}) in the main text when $T_H=1$, $R_H=0$, $|b_1|=|b_2|=1/\sqrt{2}$, and $\text{arg}\{b_1b_2^*\}=\phi_0$.
\par
Equation (\ref{signal-dm-supp}) shows that the expression of the reduced density operator representing a $d$-photon contains the state parameters, the loss parameter ($T_H$) as well as the unitary transformation parameter $\theta$. 

\subsection{Determining single-photon detection probabilities considering unequal emission probabilities and imperfect alignment}\label{app:photon-detection-probability}
We now determine the detection probabilities of a $d$-photon in all configurations.
Using Eqs.~(\ref{E-H}), (\ref{ph-countg-rt}) and (\ref{signal-dm-supp}), we find for $H$-polarization that
        \begin{align}
        &[P_H(\theta,\delta)]_j=\frac{|b_1|^2}{2}(I_{HH}+I_{VH})+\frac{|b_2|^2}{2} \cos^2 2\delta \nonumber\\& +|b_1||b_2|T_H\cos (2\delta) \Big\{\sqrt{I_{HH}}(J_{H_1H_1}^{H_2H_2})_j\cos 2\theta \nonumber\\&\times\sin\left[\phi+(\phi_{H_1H_1}^{H_2H_2})_j\right]+\sqrt{I_{VH}} (J_{V_1H_1}^{H_2H_2})_j\sin 2\theta \nonumber\\&\times\sin\left[\phi+(\phi_{V_1H_1}^{H_2H_2})_j\right]\Big\}. \label{P-H-supp}
        \end{align}
Likewise, using Eqs.~(\ref{E-V}), (\ref{ph-countg-rt}) and (\ref{signal-dm-supp}) for $V$-polarization, we obtain
        \begin{align}
        &[P_V(\theta,\delta)]_j=\frac{|b_1|^2}{2}(I_{HV}+I_{VV})+\frac{|b_2|^2}{2} \sin^2 2\delta \nonumber \\ &+|b_1||b_2|T_H \sin (2\delta) \Big\{\sqrt{I_{HV}}(J_{H_1V_1}^{H_2H_2})_j\cos 2\theta \nonumber\\&\times\sin\left[\phi+(\phi_{H_1V_1}^{H_2H_2})_j\right]+\sqrt{I_{VV}} (J_{V_1V_1}^{H_2H_2})_j\sin 2\theta\nonumber\\&\times \sin\left[\phi+(\phi_{V_1V_1}^{H_2H_2})_j\right]\Big\}\label{P-V-supp}.
        \end{align}
   Here, $j=$ A, B, C, D represents a configuration, $(J_{\mu_1\nu_1}^{H_2H_2})_j$ and $(\phi_{\mu_1\nu_1}^{H_2H_2})_j$ are given in Table~\ref{tab:1}, and  $\phi=\text{arg}\{b_1\}-\text{arg}\{b_2\}+\phi_u-\phi_d$.
We observe that Eqs.~(\ref{P-H-supp}) and (\ref{P-V-supp}) reduce to Eqs.~(\ref{P-H}) and (\ref{P-V}), respectively, in the main text when $T_H=1$, $R_H=0$, $|b_1|=|b_2|=1/\sqrt{2}$, and $\text{arg}\{b_1b_2^*\}=\phi_0$.
\par
We choose the same values of the half-wave plate angles ($\theta, \delta$) as those in the main text.  Using Eqs.~(\ref{P-H-supp}), (\ref{P-V-supp}), and Table~\ref{tab:1}, we find that the corresponding single-photon detection probabilities are given by
\begin{widetext}
    \begin{subequations}
	\begin{align}
	\text{configuration A: }&[P_H(0,0)]_{\text{A}}=\frac{|b_1|^2}{2}(I_{HH}+I_{VH})+\frac{|b_2|^2}{2} +T_H|b_1||b_2|\sqrt{I_{HH}}\sin(\phi-(\phi_p)_{\text{A}})\label{p-h-1-phase-1-supp},\\
	&[P_H(\pi/4,0)]_{\text{A}}=\frac{|b_1|^2}{2}(I_{HH}+I_{VH})+\frac{|b_2|^2}{2} +T_H|b_1||b_2|\sqrt{I_{VH}}J_{VH}^{HH}\sin(\phi-(\phi_p)_{\text{A}}+\phi_{VH}^{HH})\label{p-h-1-phase-2-supp},\\
	&[P_V(0,\pi/4)]_{\text{A}}=\frac{|b_1|^2}{2}(I_{HV}+I_{VV})+\frac{|b_2|^2}{2} +T_H|b_1||b_2|\sqrt{I_{HV}}J_{HV}^{HH}\sin(\phi-(\phi_p)_{\text{A}}+\phi_{HV}^{HH})\label{p-v-1-phase-1-supp},\\
	&[P_V(\pi/4,\pi/4)]_{\text{A}}=\frac{|b_1|^2}{2}(I_{HV}+I_{VV})+\frac{|b_2|^2}{2} +T_H|b_1||b_2|\sqrt{I_{VV}}J_{VV}^{HH}\sin(\phi-(\phi_p)_{\text{A}}+\phi_{VV}^{HH})\label{p-v-1-phase-2-supp},\\
    \text{configuration B: }&[P_H(0,0)]_{\text{B}}=\frac{|b_1|^2}{2}(I_{HH}+I_{VH})+\frac{|b_2|^2}{2}+T_H|b_1||b_2|\sqrt{I_{HH}}J_{HH}^{HV}\sin(\phi-(\phi_p)_{\text{B}}+\phi^{HV}_{HH})\label{p-h-2-phase-1-supp},\\
	&[P_H(\pi/4,0)]_{\text{B}}=\frac{|b_1|^2}{2}(I_{HH}+I_{VH})+\frac{|b_2|^2}{2} +T_H|b_1||b_2|\sqrt{I_{VH}}J_{VH}^{HV}\sin(\phi-(\phi_p)_{\text{B}}+\phi_{VH}^{HV})\label{p-h-2-phase-2-supp},\\
	&[P_V(0,\pi/4)]_{\text{B}}=\frac{|b_1|^2}{2}(I_{HV}+I_{VV})+\frac{|b_2|^2}{2} +T_H|b_1||b_2|\sqrt{I_{HV}}\sin(\phi-(\phi_p)_{\text{B}})\label{p-v-2-phase-1-supp},\\
	&[P_V(\pi/4,\pi/4)]_{\text{B}}=\frac{|b_1|^2}{2}(I_{HV}+I_{VV})+\frac{|b_2|^2}{2} +T_H|b_1||b_2|\sqrt{I_{VV}}J_{VV}^{HV}\sin(\phi-(\phi_p)_{\text{B}}+\phi_{VV}^{HV})\label{p-v-2-phase-2-supp},\displaybreak\\
    \text{configuration C: }&[P_H(0,0)]_{\text{C}}=\frac{|b_1|^2}{2}(I_{HH}+I_{VH})+\frac{|b_2|^2}{2} +T_H|b_1||b_2|\sqrt{I_{HH}}J_{HH}^{VH}\sin(\phi-(\phi_p)_{\text{C}}+\phi^{VH}_{HH})\label{p-h-3-phase-1-supp},\\
    &[P_H(\pi/4,0)]_{\text{C}}=\frac{|b_1|^2}{2}(I_{HH}+I_{VH})+\frac{|b_2|^2}{2} +T_H|b_1||b_2|\sqrt{I_{VH}}\sin(\phi-(\phi_p)_{\text{C}})\label{p-h-3-phase-2-supp},\\
    &[P_V(0,\pi/4)]_{\text{C}}=\frac{|b_1|^2}{2}(I_{HV}+I_{VV})+\frac{|b_2|^2}{2} +T_H|b_1||b_2|\sqrt{I_{HV}}J_{HV}^{VH}\sin(\phi-(\phi_p)_{\text{C}}+\phi_{HV}^{VH})\label{p-v-3-phase-1-supp},\\
    &[P_V(\pi/4,\pi/4)]_{\text{C}}=\frac{|b_1|^2}{2}(I_{HV}+I_{VV})+\frac{|b_2|^2}{2} +T_H|b_1||b_2|\sqrt{I_{VV}}J_{VV}^{VH}\sin(\phi-(\phi_p)_{\text{C}}+\phi_{VV}^{VH})\label{p-v-3-phase-2-supp},\\
    \text{configuration D: }&[P_H(0,0)]_{\text{D}}=\frac{|b_1|^2}{2}(I_{HH}+I_{VH})+\frac{|b_2|^2}{2} +T_H|b_1||b_2|\sqrt{I_{HH}}J_{HH}^{VV}\sin(\phi-(\phi_p)_{\text{D}}+\phi^{VV}_{HH})\label{p-h-4-phase-1-supp},\\
    &[P_H(\pi/4,0)]_{\text{D}}=\frac{|b_1|^2}{2}(I_{HH}+I_{VH})+\frac{|b_2|^2}{2} +T_H|b_1||b_2|\sqrt{I_{VH}}J_{VH}^{VV}\sin(\phi-(\phi_p)_{\text{D}}+\phi^{VV}_{VH})\label{p-h-4-phase-2-supp},\\
    &[P_V(0,\pi/4)]_{\text{D}}=\frac{|b_1|^2}{2}(I_{HV}+I_{VV})+\frac{|b_2|^2}{2} +T_H|b_1||b_2|\sqrt{I_{HV}}J_{HV}^{VV}\sin(\phi-(\phi_p)_{\text{D}}+\phi_{HV}^{VV})\label{p-v-4-phase-1-supp},\\
    &[P_V(\pi/4,\pi/4)]_{\text{D}}=\frac{|b_1|^2}{2}(I_{HV}+I_{VV})+\frac{|b_2|^2}{2} +T_H|b_1||b_2|\sqrt{I_{VV}}\sin(\phi-(\phi_p)_{\text{D}})\label{p-v-4-phase-2-supp},
	\end{align}
\end{subequations}
\end{widetext}
where $\phi=\text{arg}\{b_1\}-\text{arg}\{b_2\}+\phi_u-\phi_d$.
\par
The detection probabilities given by Eqs.~(\ref{p-h-1-phase-1-supp})-(\ref{p-v-4-phase-2-supp}) represent single-photon interference patterns. The visibility of these interference patterns is given by the standard formula
\begin{align}\label{vis-def}
    \left[\mathcal{V}_\mu(\theta,\delta)\right]_j=\frac{\{[P_\mu(\theta,\delta)]_j\}_{\text{max}}-\{[P_\mu(\theta,\delta)]_j\}_{\text{min}}}{\{[P_\mu(\theta,\delta)]_j\}_{\text{max}}+\{[P_\mu(\theta,\delta)]_j\}_{\text{min}}},
\end{align}
where maximum (max) and minimum (min) values of $P_\mu(\theta,\delta)$ are obtained by varying $\phi$.
\subsection{Determining $|b_1|$, $|b_2|$, and $T_H$ in terms of experimentally measurable quantities}\label{app:loss-characterize}
We first show how to determine $|b_1|$ and $|b_2|$ from the measurable photon detection probabilities (photon counting rates).
Setting $\delta=\pi/4$ in Eq.~(\ref{P-H-supp}) and $\delta=0$ in Eq.~(\ref{P-V-supp}), we find that
\begin{subequations}
   \begin{align}
    &[P_H(\theta,\pi/4)]_j=\frac{|b_1|^2}{2}(I_{HH}+I_{VH})\equiv P_H^{(0)},\label{P-b1-H-supp}\\
     &[P_V(\theta,0)]_j=\frac{|b_1|^2}{2}(I_{HV}+I_{VV})\equiv P_V^{(0)},\label{P-b1-V-supp}
\end{align} 
\end{subequations}
where both $P_H^{(0)}$ and $P_V^{(0)}$ are independent of $\theta$ and remains the same in all configurations. 
Since $\sum_{\mu,\nu}^{H,V}I_{\mu\nu}=1$, it follows from Eqs.~(\ref{P-b1-H-supp}) and (\ref{P-b1-V-supp}) that
\begin{subequations}\label{b1b2-supp}
        \begin{align}
             &|b_1|^2=2\left[P_H^{(0)}+P_V^{(0)}\right], \label{b1-supp}\\
             &|b_2|^2=1-2\left[P_H^{(0)}+P_V^{(0)}\right] \label{b2-supp}.
        \end{align}
\end{subequations}
\par
We now show how to determine $T_H$, that characterizes dominant experimental imperfections. 
Using Eqs.~(\ref{vis-def}), (\ref{p-h-1-phase-1-supp}), and (\ref{p-h-3-phase-2-supp}), we obtain
\begin{subequations}
    \begin{align}
        &[\V_H(0,0)]_{\text{A}}=\frac{2T_H|b_1||b_2|\sqrt{I_{HH}}}{|b_1|^2(I_{HH}+I_{VH})+|b_2|^2},\label{v-h-a-1-supp}\\
        &[\V_H(\pi/4,0)]_{\text{C}}=\frac{2T_H|b_1||b_2|\sqrt{I_{VH}}}{|b_1|^2(I_{HH}+I_{VH})+|b_2|^2}. \label{v-h-c-2-supp}
    \end{align}
\end{subequations}
It now follows from Eqs.~(\ref{P-b1-H-supp})-(\ref{v-h-c-2-supp}) that
\begin{align}\label{vis-combo}
        ([\V_H(0,0)]_{\text{A}})^2+([\V_H(\pi/4,0)]_{\text{C}})^2=\frac{8T_H^2|b_2|^2P_H^{(0)}}{(1-2P_V^{(0)})^2}.
    \end{align}
Finally, using Eqs.~(\ref{b2-supp}) and (\ref{vis-combo}) we find that
\begin{align}\label{t-h-2}
    (T_H)^2&=\big\{([\V_H(0,0)]_{\text{A}})^2+([\V_H(\pi/4,0)]_{\text{C}})^2\big\}\nonumber\\&\times\frac{\left(1-2P_V^{(0)}\right)^2}{8P_H^{(0)}\big(1-2[P_H^{(0)}+P_V^{(0)}]\big)}.
\end{align}
We have thus represented $T_H$ in terms of single-photon counting data.
\subsection{Reconstruction of the density matrix considering unequal emission probabilities and imperfect alignment}\label{app:state-reconstruction-loss}
We now show how to reconstruct the density matrix of the unknown quantum state $\dm$ [Eq.~(\ref{q-state})] when the sources emit with unequal probabilities and the alignment of undetected photon beams is imperfect.
\par
Using Eqs.~(\ref{q-state}) and 
 (\ref{Ppm}) from the main text and Eqs.~(\ref{p-h-1-phase-1-supp})-(\ref{p-v-4-phase-2-supp}), we find that 
\begin{widetext}
    \begin{subequations}
    \begin{align}
        &\bra{H_u H_d} \,\dm \,\ket{H_u H_d}=I_{HH}=\Gamma\left\{[P_H^{(-)}(0,0)]_A\right\}^2\label{I-11-supp},\\
        &\bra{H_u V_d} \,\dm \,\ket{H_u V_d}=I_{HV}=\Gamma\left\{[P_V^{(-)}(0,\pi/4)]_B\right\}^2\label{I-22-supp},\\
        &\bra{V_u H_d} \,\dm \,\ket{V_u H_d}=I_{VH}=\Gamma\left\{[P_H^{(-)}(\pi/4,0)]_C\right\}^2\label{I-33-supp},\\
        &\bra{V_u V_d} \,\dm \,\ket{V_u V_d}=I_{VV}=\Gamma\left\{[P_V^{(-)}(\pi/4,\pi/4)]_D\right\}^2\label{I-44-supp},\\
        &|\bra{H_u H_d} \,\dm \,\ket{H_u V_d}|=\sqrt{I_{HH}I_{HV}}J_{HV}^{HH}=\Gamma[P_H^{(-)}(0,0)]_A [P_V^{(-)}(0,\pi/4)]_A=\Gamma [P_V^{(-)}(0,\pi/4)]_B[P_H^{(-)}(0,0)]_B\label{d-12-supp},\\
        &|\bra{H_u H_d} \,\dm \,\ket{V_u H_d}|=\sqrt{I_{HH}I_{VH}}J_{VH}^{HH}=\Gamma[P_H^{(-)}(0,0)]_A [P_H^{(-)}(\pi/4,0)]_A=\Gamma[P_H^{(-)}(\pi/4,0)]_C[P_H^{(-)}(0,0)]_C\label{d-13-supp},\\
        &|\bra{H_u H_d} \,\dm \,\ket{V_u V_d}|=\sqrt{I_{HH}I_{VV}}J_{VV}^{HH}=\Gamma[P_H^{(-)}(0,0)]_A [P_V^{(-)}(\pi/4,\pi/4)]_A=\Gamma[P_V^{(-)}(\pi/4,\pi/4)]_D[P_H^{(-)}(0,0)]_D\label{d-14-supp},\\
        &|\bra{H_u V_d} \,\dm \,\ket{V_u H_d}|=\sqrt{I_{HV}I_{VH}}J_{VH}^{HV}=\Gamma[P_V^{(-)}(0,\pi/4)]_B[P_H^{(-)}(\pi/4,0)]_B=\Gamma[P_H^{(-)}(\pi/4,0)]_C[P_V^{(-)}(0,\pi/4)]_C\label{d-23-supp},\\
        &|\bra{H_u V_d} \,\dm \,\ket{V_u V_d}|=\sqrt{I_{HV}I_{VV}}J_{VV}^{HV}=\Gamma[P_V^{(-)}(0,\pi/4)]_B[P_V^{(-)}(\pi/4,\pi/4)]_B=\Gamma[P_V^{(-)}(\pi/4,\pi/4)]_D[P_V^{(-)}(0,\pi/4)]_D\label{d-24-supp},\\
        &|\bra{V_u H_d} \,\dm \,\ket{V_u V_d}|=\sqrt{I_{VH}I_{VV}}J_{VV}^{VH}=\Gamma[P_H^{(-)}(\pi/4,0)]_C[P_V^{(-)}(\pi/4,\pi/4)]_C=\Gamma[P_V^{(-)}(\pi/4,\pi/4)]_D[P_H^{(-)}(\pi/4,0)]_D\label{d-34-supp},
    \end{align}
\end{subequations}
\end{widetext}
where $\Gamma$ is an experimentally measurable quantity given by $\Gamma=1/\left(4T_H^2|b_1|^2|b_2|^2\right)$. It is evident that $\Gamma$ can be expressed in terms of single-photon counting data by using expressions for $|b_1|$, $|b_2|$, and $T_H$ from Eqs.~(\ref{b1-supp}), (\ref{b2-supp}), and (\ref{t-h-2}), respectively. We find that
\begin{align}
    \Gamma=&\frac{1}{\big\{([\V_H(0,0)]_{\text{A}})^2+([\V_H(\pi/4,0)]_{\text{C}})^2\big\}}\nonumber\\&\times\frac{P_H^{(0)}}{(1-2P_V^{(0)})^2\left[P_H^{(0)}+P_V^{(0)}\right]}.
\end{align}
It can be readily checked that when $T_H=1$ and $|b_1|=|b_2|=1/\sqrt{2}$, $\Gamma$ becomes equal to $1$.
\par
We have thus determined all diagonal elements [Eqs.~(\ref{I-11-supp})-(\ref{I-44-supp})] and moduli of all off-diagonal elements [Eqs.~(\ref{d-12-supp})-(\ref{d-34-supp})] of $\dm$ considering unequal emission probabilities and imperfect alignment of undetected photon beams.
\par
The phases appearing in the expressions of the detection probabilities given in Eqs.~(\ref{p-h-1-phase-1-supp})-(\ref{p-v-4-phase-2-supp}) are same as those in the lossless scenario [Eqs.~(\ref{p-h-1-phase-1})-(\ref{p-v-4-phase-2}) in the main text]. Therefore, the method described in Sec.~\ref{subsec:state-reconstruction} to determine the arguments of the off-diagonal matrix elements directly applies when experimental imperfections are present. 

\section{Graphs for configurations B, C and D } \label{app:q-state-full}
\begin{figure}[htbp]    \includegraphics[width=\linewidth]{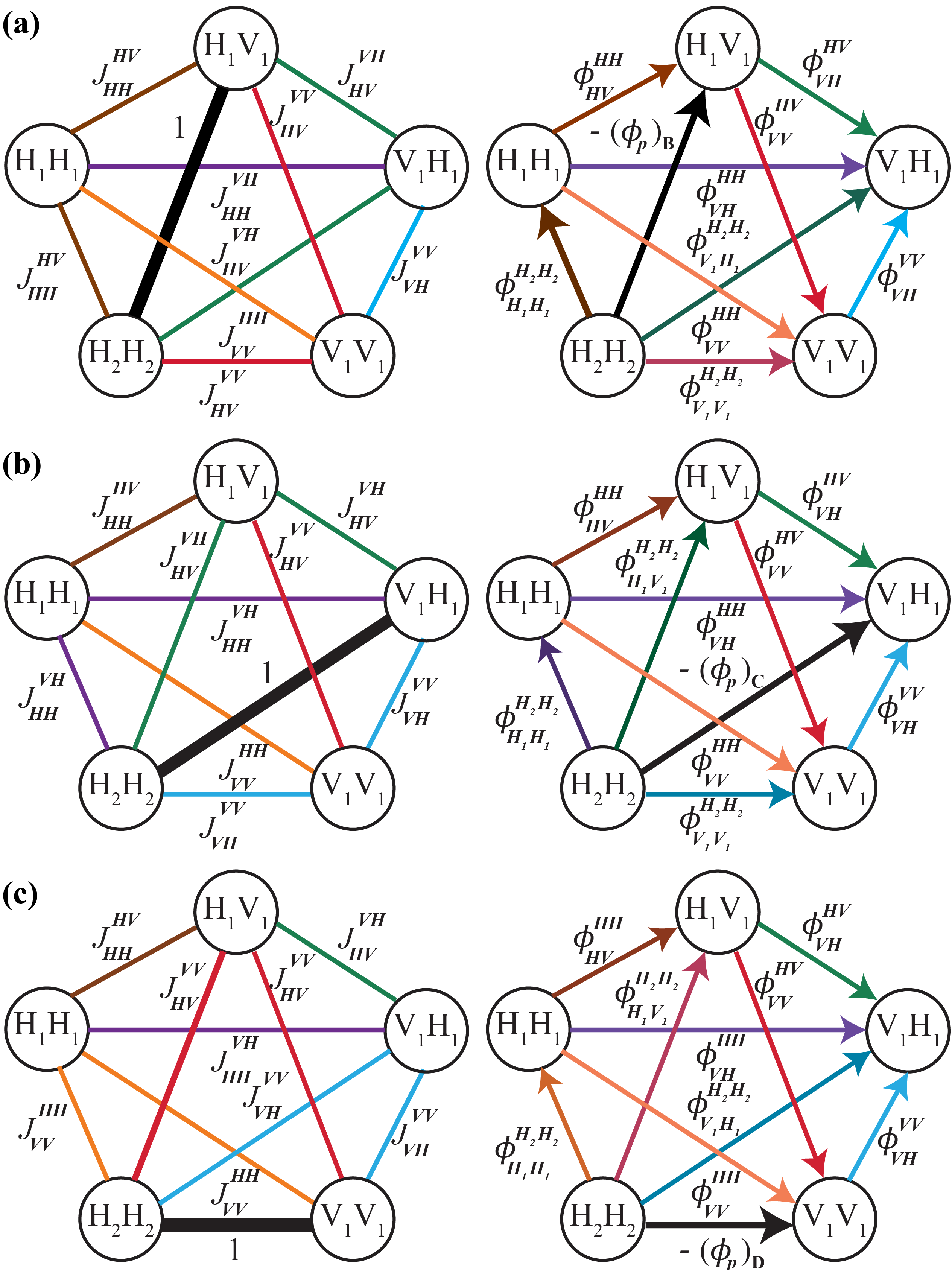}
    \caption{Graphs representing the quantum state in configurations B, C, and D before path identity is employed are shown in figures (a), (b), and (c), respectively. (Intensity parameters are not shown.)}\label{config-B,C,D-para-graphs}
\end{figure}
The graphs corresponding to indistinguishability and phase parameters (Table~\ref{tab:1}) in configurations B, C, and D are given in Fig.~\ref{config-B,C,D-para-graphs}.

\section{Choosing configurations without prior information}\label{app:finding-configs}
We now show that one does not need any prior information about the quantum state to identify the four configurations (A, B, C, D) of the setup (Figs.~\ref{fig:setup} and \ref{fig:scheme-supp}). 
\par
Equations (\ref{P-H-supp}) and (\ref{P-V-supp}) show that the interference patterns can be modulated by varying the experimentally controllable parameters (half-wave plate angles) $\delta$ and $\theta$. 
Using Eqs.~(\ref{P-H-supp}) and (\ref{vis-def}), and setting $\delta=0$ and $\theta=0$, we obtain the visibility corresponding to $H$-polarization as
\begin{align}\label{v-h-a-part-1}
    [\V_H(0,0)]_j=\frac{2|b_1||b_2|T_H\sqrt{I_{HH}}(J_{H_1H_1}^{H_2H_2})_j}{|b_1|^2(I_{HH}+I_{VH})+|b_2|^2},
\end{align}
where $j=\text{A},\text{B},\text{C},\text{D}$ represent a configuration.
We observe that for each configuration $[V_H(0,0)]_j\propto (J_{H_1H_1}^{H_2H_2})_j$. Values of $(J_{H_1H_1}^{H_2H_2})_j$ for configurations $j = $ A, B, C, and D are 1, $J_{HH}^{HV}$, $J_{HH}^{VH}$, and $J_{HH}^{VV}$, respectively (Table~\ref{tab:1} in main text). Since $0\leq J_{\mu\nu}^{\mu'\nu'}\leq 1$, we find that the visibility $[V_H(0,0)]_j$ attains its maximum value in configuration A, i.e.,
\begin{align}
    \max_j [V_H(0,0)]_j=[V_H(0,0)]_A.
\end{align}
Therefore, configuration A can be reached by varying the tunable delay (Figs.~\ref{fig:setup} and \ref{fig:scheme-supp}) until $V_H(0,0)$ attains the maximum value. 
\par
Likewise, using Eq.~(\ref{P-H-supp}), Eq.~(\ref{P-V-supp}), Eq.~(\ref{vis-def}), and Table~\ref{tab:1}, we find that
	\begin{subequations}
		\begin{align}
        &\max_j[\V_V(0,\pi/4)]_j=\frac{2T_H|b_1||b_2|\sqrt{I_{HV}}}{|b_1|^2(I_{HV}+I_{VV})+|b_2|^2}\nonumber\\&= [\V_V(0,\pi/4)]_{\text{B}}\label{v-v-1},\\
		&\max_j[\V_H(\pi/4,0)]_j=\frac{2T_H|b_1||b_2|\sqrt{I_{VH}}}{|b_1|^2(I_{HH}+I_{VH})+|b_2|^2}\nonumber\\&=[\V_H(\pi/4,0)]_{\text{C}}\label{v-h-2},\\
		&\max_j[\V_V(\pi/4,\pi,4)]_j=\frac{2T_H|b_1||b_2|\sqrt{I_{VV}}}{|b_1|^2(I_{HV}+I_{VV})+|b_2|^2}\nonumber\\&= [\V_V(\pi/4,\pi,4)]_{\text{D}}\label{v-v-2}.
		\end{align}
	\end{subequations}
Therefore, configurations B, C, and D are identified by setting $\V_V(0,\pi/4)$, $\V_H(\pi/4,0)$, and $\V_V(\pi/4,\pi,4)$, respectively, to their maximum values by appropriately adjusting the tunable optical delay between the pump beams. 

\par
If any of the visibilities given by Eqs.~(\ref{v-h-a-part-1})-(\ref{v-v-2}) is identically zero, the corresponding configuration does not exist. For example, in the special case illustrated in Sec.~\ref{sec:illustration}, $\V_V(0,\pi/4)$ and $\V_H(\pi/4,0)$ can be shown to be identically zero, and therefore, configurations B and C do not exist.
\par
If the unknown quantum state is pure, i.e., if $J_{\mu\nu}^{\mu'\nu'}=1$ for all $\mu$, $\nu$, $\mu'$, and $\nu'$, then the four configurations are indistinguishable from each other. In this case, all state parameters can be determined by using a single configuration. We also note that a single two-qubit pure state can only contain up to nine free parameters instead of fifteen.

\bibliography{tomo-PI-sp}

\begin{thebibliography}{43}
\bibitem[{ref()}]{ref-state-note}
\bibinfo{note}{Any other basic state such as
  $\ket{\psi_{ref}}=\ket{V_{u_2}V_{d_2}}$, or
  $\ket{\psi_{ref}}=\ket{H_{u_2}V_{d_2}}$, or
  $\ket{\psi_{ref}}=\ket{V_{u_2}H_{d_2}}$ can also be chosen.}
\bibitem[{non()}]{nonzero-vis}
\bibinfo{note}{If, due to experimental imperfections, interference patterns
  with very low visibility are obtained in configurations B and C, one needs to
  use Eqs.~(\ref{I-11})-(\ref{d-34}) and Table~\ref{phase-tab-res}.}
\end{thebibliography}

\end{document}